\documentclass[10pt,journal,compsoc]{IEEEtran}

\ifCLASSOPTIONcompsoc
  \usepackage[nocompress]{cite}
\else
  \usepackage{cite}
\fi

\usepackage{subfigure}
\usepackage{algorithmic}
\usepackage[ruled,linesnumbered]{algorithm2e}

\usepackage{amsmath,amssymb,amsfonts}
\usepackage{graphicx}
\usepackage{color}
\usepackage{ragged2e}
\usepackage{enumerate}

\ifCLASSINFOpdf
\else
\fi

\makeatletter
\newcommand{\removelatexerror}{\let\@latex@error\@gobble}
\makeatother

\begin{document}
%
\title{PET: Multi-agent Independent PPO-based Automatic ECN Tuning for High-Speed \\Data Center Networks}

\author{Kai~Cheng,~\IEEEmembership{Student Member,~IEEE,}
        Ting~Wang,~\IEEEmembership{Senior Member,~IEEE,}\\
        Xiao~Du,~\IEEEmembership{Student Member,~IEEE,} 
        Shuyi Du, 
        and Haibin Cai
        \thanks{\textit{(Corresponding author: Ting Wang.)}}

\IEEEcompsocitemizethanks{
    \IEEEcompsocthanksitem This work was partially supported by the National Key Research and Development Program of China (No. 2021ZD0114600) and the grants from Shenzhen Science and Technology Plan Project (No. CJGJZD20210408092400001).
    \IEEEcompsocthanksitem Kai Cheng, Ting Wang, Xiao Du, and Haibin Cai are with the Engineering Research Center of Software/Hardware Co-design Technology and Application, Ministry of Education; the Shanghai Key Laboratory of Trustworthy Computing; Software Engineering Institute, East China Normal University, Shanghai 200050, China. (Email: 51215902085@stu.ecnu.edu.cn, twang@sei.ecnu.edu.cn, 71194501039@stu.ecnu.edu.cn, hbcai@sei.ecnu.edu.cn).  \protect\\
    \IEEEcompsocthanksitem Shuyi Du is with the Department of Houston International Institute, Dalian Maritime University, Dalian 116026, Liaoning, China. (Email: dushuyi0711@dlmu.edu.cn.)
}
}

\IEEEtitleabstractindextext{%
\begin{abstract}
\justifying
Explicit Congestion Notification (ECN)-based congestion control schemes have been widely adopted in high-speed data center networks (DCNs), where the ECN marking threshold plays a determinant role in guaranteeing a packet lossless DCN. However, existing approaches either employ static settings with immutable thresholds that cannot be dynamically self-adjusted to adapt to network dynamics, or fail to take into account many-to-one traffic patterns and different requirements of different types of traffic, resulting in relatively poor performance. To address these problems, this paper proposes a novel learning-based automatic ECN tuning scheme, named PET, based on the multi-agent Independent Proximal Policy Optimization (IPPO) algorithm. PET dynamically adjusts ECN thresholds by fully considering pivotal congestion-contributing factors, including queue length, output data rate, output rate of ECN-marked packets, current ECN threshold, the extent of incast, and the ratio of mice and elephant flows. PET adopts the Decentralized Training and Decentralized Execution (DTDE) paradigm and combines offline and online training to accommodate network dynamics. PET is also fair and readily deployable with commodity hardware. Comprehensive experimental results demonstrate that, compared with state-of-the-art static schemes and the learning-based automatic scheme, our PET achieves better performance in terms of flow completion time, convergence rate, queue length variance, and system robustness.

\end{abstract}

\begin{IEEEkeywords}
Congestion Control, Data Center Network, ECN.
\end{IEEEkeywords}}

\maketitle

\IEEEdisplaynontitleabstractindextext

%
\IEEEpeerreviewmaketitle

\ifCLASSOPTIONcompsoc
\IEEEraisesectionheading{\section{Introduction}\label{sec:introduction}}
\else
\section{Introduction}
\label{intro}
\fi

%
%
%
%
\IEEEPARstart{W}{ith} the advent of the cloud era, computation and storage have progressively shifted to the cloud. As the key infrastructure of cloud computing, the data center (DC) composed of an enormous number of servers and network devices has played an increasingly critical role in supporting the powerful computing and massive storage needs of individuals and enterprises.
To meet the ever-growing demands of cloud-based services, the data center is experiencing a dramatic increase in the number of servers, which in turn requires a vast number of network devices to form an interconnected system, resulting in a rapidly expanding network scale with high networking complexity.
Thus, how to provide effective congestion control (CC) in such a complex, dynamic and diversified large-scale data center network (DCN) to ensure a high-quality and responsive network service is confronted with multi-faceted challenges, which mainly lie in three aspects.

On the one hand, modern cloud data centers are usually equipped with a large collection of computation- or data-intensive applications, such as complex image processing, scientific computing, big data processing \cite{chen2015mxnet,lu2017octopus,abadi2016tensorflow,wei2018deconstructing}, distributed storage \cite{barthels2015rack,lu2016high,wang2015hydradb}, and artificial intelligence (AI) model training \cite{9206551}, which have thereby spawned many distributed computing frameworks, like MapReduce \cite{dean2008mapreduce}, Spark \cite{zaharia2010spark}, and Flink \cite{carbone2015apache}, aiming to deliver high performance computing \cite{lu2013high,lu2014accelerating,zhang2017performance}. However, such distributed computing paradigms continuously generate a large amount of many-to-one partition-aggregate traffic patterns with high fan-in, which inevitably results in intractable incast issues accompanied by persistent queue build-up, increased delay, jitter, and even packet loss \cite{li2022rethinking}. Thus, how to design an incast-aware congestion control scheme becomes an imperative concern for the high-speed DCN.
On the other hand, as a diversified environment, cloud data centers typically provide a variety of services that produce various types of traffic with different characteristics expecting different requirements for network quality. For instance, large long-running elephant flows (e.g. data replication, virtual machine migration) usually entail high requirements for throughput, which can be preferably achieved by long queue settings on the switch side but have a certain tolerance for network latency.
Comparatively, small short-lived mice flows (e.g. control, management, and query messages) impose rigorous restrictions on packet delay, which prefers short queue length on the switch side, but rarely has requirements for throughput. 
As a result, how to adaptively adjust the queue length to simultaneously meet these conflicting requirements of different types of traffics remains another key challenge.
Last but not least, the data center network is recognized as a highly dynamic network environment, where the traffic amount, traffic patterns, and the ratio between mice and elephant flows are changing rapidly, which brings significant uncertainty to CC mechanisms. 
This thereby raises another critical challenge, that is, how to enable the network congestion control strategy with the ability of self-learning and self-decision-making  to dynamically adapt to real-time network conditions.


Explicit Congestion Notification (ECN)  has been acknowledged as an effective means to facilitate network congestion control and has been widely supported by commodity switches in data centers \cite{li2019hpcc}. 
Among these existing ECN-based CC schemes, the setting strategy of the ECN marking threshold plays a vital role in determining their feasibility and effectiveness. Generally, there are mainly three strategies to set the ECN marking threshold, i.e., static settings \cite{alizadeh2010data,vamanan2012deadline,9155280,xia2021multipath,hu2021aeolus,8372948}, dynamic settings \cite{zhang2019enabling,ZHANG2016222,LU2018197,8791847}, and learning-based automatic settings \cite{yan2021acc}. 
The static schemes require the switches to be pre-configured with a fixed ECN marking threshold throughout the execution cycle of the algorithm. However, such static settings evidently can neither adapt to network dynamics nor can simultaneously meet different requirements of mice and elephant flows, where a high threshold can lead to missed deadlines for delay-sensitive mice flows while a low threshold can result in throughput degradation for bandwidth-hungry elephant flows.
Comparatively, dynamic schemes can adjust the ECN marking threshold based on some simplistic judgment mechanisms in a dynamic manner. 
However, the adjustment policies should be manually pre-defined and cannot be self-adjusted according to the real-time network conditions. Worse still, existing dynamic methods either consider only one single simple factor (such as link utilization \cite{ZHANG2016222} or instantaneous queue length \cite{8791847})  to adjust the threshold, with limited performance, or only apply to multi-queues and cannot be used for a single queue \cite{LU2018197}\cite{MAJIDI2020334}.
Reinforcement learning (RL), which can enable agents to dynamically make decisions with the maximum rewards through continuously interacting with the environment, provides an effective way to deal with the above issues.
Nevertheless, the existing RL-based ECN tuning schemes are relatively few. To the best of our knowledge, ACC \cite{yan2021acc} is the first and only RL-based automatic ECN tuning method for DCN, which generates appropriate policies based on the observed statistics and updates the ECN threshold through the switch’s control interface. Experimental results prove the effectiveness of ACC, which significantly outperforms both static and dynamic schemes.
However, ACC also has its notable deficiencies, which seriously inhibit its performance. First of all, ACC only considers some basic metrics (such as current queue length, data rate, and current ECN threshold) in learning threshold adjustment strategies based on its deep reinforcement learning (DRL) model, failing to take into account other equally important metrics including the extent of incast and the ratio of mice and elephant flows. This will inevitably lead to that the model cannot fully understand the network states, thus making the learned policies cannot always be optimal,
especially in the presence of incast and mice-elephant mixed flows. Secondly, ACC directly adopts the simple Double Deep Q-Network (DDQN) algorithm \cite{van2016deep} as its learning model.  However, DDQN requires global experience replay in a multi-agent scenario, which leads to a certain degree of memory overhead and bandwidth consumption. This deems to be impractical and unacceptable  for resource-constrained switches.


To address the above-mentioned issues, this paper proposes a novel multi-agent reinforcement learning (MARL)-based automatic ECN tuning approach, named PET, based on the state-of-the-art multi-agent actor-critic Independent Proximal Policy Optimization (IPPO) algorithm \cite{de2020independent}. 
PET is implemented based on the Decentralized Training with Decentralized Execution (DTDE) framework, which combines offline training and online incremental training to dynamically adjust the ECN marking threshold according to real-time network states without manual intervention. 
To achieve better adaptability to network dynamics with a more comprehensive knowledge of network states, our PET incorporates six pivotal metrics, i.e., queue length, the output data rate for each link, the extent of incast, the ratio of mice and elephant flows, the output rate of ECN marked packets, and current ECN threshold, into the input of our MARL model, aiming to learn a more effective ECN tuning policy. 
Different from ACC which applies the DDQN algorithm, our PET employs the latest and more effective IPPO algorithm integrated with our modified reward function as the learning algorithm. 
Note that, unlike DDQN, IPPO does not require the global experience replay in the multi-agent scenario, which can effectively avoid unnecessary memory overhead and bandwidth consumption.
Comprehensive experimental results demonstrate that, compared with static ECN schemes and the state-of-the-art learning-based ACC approach, PET can achieve more reduction in flow completion time (FCT) for mice flows in terms of both the average FCT and 99-th FCT, while without affecting the throughput of elephant flows.
Our large-scale simulations also prove that, with our PET approach, it is possible to achieve an excellent balance between high throughput, low latency, and high link utilization. Importantly, PET is well-compatible with existing ECN-based schemes and requires no modifications to the ECN-based rate control on the server side.
Besides, PET also can be easily integrated into the commodity switches requiring no special features of switching chips. 
To summarize, this paper makes the following three major contributions:
\begin{itemize}
\item We propose a novel learning-based automatic ECN tuning algorithm, named PET, based on the latest MARL algorithm IPPO, working in a distributed manner. PET can implement a ``zero-touch" automatic ECN tuning in response to network dynamics without manual configurations, and it can be easily deployed with good compatibility with existing ECN-based schemes. 

\item We design a more reasonable and comprehensive ECN threshold quantification mechanism with better practicability by taking into account six pivotal factors contributing to congestion, including the extent of incast and the ratio of mice and elephant flows, enabling the DRL agents to have a better understanding of the network conditions and thus make more accurate ECN configurations to achieve better performance. To the best of our knowledge, this is the first attempt in this field.

\item We improve the IPPO model with a modified reward function to accommodate the optimization goals, which can expedite the convergence speed and improve the robustness of the DRL algorithm.

\end{itemize}
We conduct extensive simulations to validate the effectiveness of PET and compare it with both static schemes and the state-of-the-art learning-based scheme. Experimental results demonstrate that PET achieves better performance in various aspects.

The rest of the paper is organized as follows. 
Section \ref{relatedwork} briefly reviews the related works of ECN tuning schemes. Section \ref{motivation} provides some preliminaries. Section \ref{approach} details the design of our PET approach. Section \ref{evaluation} presents the evaluation results. Finally, Section \ref{conclusion} concludes the paper.

\section{Related Work}
\label{relatedwork}



In view of the effectiveness of ECN in improving the delay and throughput of high-speed DCNs, 
the research on ECN-based DCN congestion control has attracted increasing attention, and a large number of ECN-based CC schemes have been proposed.
Generally, existing ECN-based schemes can be classified into three categories, i.e., static ECN schemes, dynamic ECN tuning schemes, and learning-based automatic ECN tuning schemes.
\subsection{Static ECN Schemes}

Static ECN schemes refer to that the ECN marking threshold is pre-configured and remains immutable without any adjustments during the whole process of the algorithm. 
DCTCP \cite{alizadeh2010data} is a pioneer DCN CC scheme based on static ECN settings. In DCTCP, 
servers adjust their sending rate according to the proportion of packets marked with ECN within the current window.
In this way, DCN can ensure a queue with a short average length, which can help absorb burst traffic and alleviate incast problems to a certain extent. However, DCTCP is a deadline agnostic protocol, where around 25\% of deadlines will be missed at high fan-in with tight deadlines \cite{vamanan2012deadline}.
To this end, based on DCTCP, D$^2$TCP \cite{vamanan2012deadline} additionally takes into account the deadline imminence factor when adjusting the congestion window size at the server side.
However, both DCTCP and D$^2$TCP do not consider the fairness problem, equally treating all traffic without priority settings.
DCQCN \cite{zhu2015congestion} provides an end-to-end rate-based CC method based on the priority-based flow control (PFC) at the ingress queue together with the ECN marking at the egress queue of a switch.
CEDM \cite{8372948} investigates the buffer underflow problem caused by the instantaneous-queue-length-based ECN marking due to over-tuning, and proposes a new marking method by considering the length variation rate of both the enqueue and dequeue to reduce queue oscillations.


BCC \cite{9155280} adopts a per-port ECN/RED (inherited from DCTCP) scheme for the case of sufficient buffer space and a shared buffer ECN/RED scheme for the case of insufficient buffer space to work together, aiming to achieve a packet-lossless DCN.
\cite{hu2021aeolus} implements a RED/ECN-based selective packet dropping method using the common static ECN settings to differently treat the pre-configured unscheduled packets and scheduled packets.


\subsection{Dynamic ECN Tuning Schemes}






Since static ECN tuning mechanisms cannot adapt to network dynamics, researchers began to shift their attention to dynamic ECN tuning schemes, in which the ECN marking threshold can be dynamically adjusted following a pre-defined rule.
AMT \cite {ZHANG2016222} adjusts the ECN marking threshold according to the link utilization, which is periodically detected by switches. 
LU et al. \cite{LU2018197}  propose two dynamic ECN tuning schemes called DEMT and EDEMT, which respectively set different thresholds for large flow queues and small flow queues to accommodate different flows' requirements.
QAECN \cite{8791847} also targets multi-queue switches and sets the ECN threshold for each queue based solely on the instantaneous queue length.
DC-ECN \cite{MAJIDI2020334} employs a machine learning method to separate the mice and elephant flows, which are then injected into respective queues. In case the mice flow queue exceeds the ECN threshold, its packets will be transferred to the elephant flow queue and sent with the highest priority so as to guarantee the low latency of mice flows. However, this practice may lead to starving elephant flows. 
ECN$^\sharp$ \cite{zhang2019enabling} designs an ECN marking approach based on both instantaneous queue length and sojourn time, which enables ECN to deal with RTT variations.
Although dynamic ECN schemes can alleviate the problems of static schemes, they heavily rely on the manually pre-defined and pre-configured adjustment rules, which usually only consider one or two simple factors making the schemes inherently still not well adaptive to the network dynamics with limited performance.

\subsection{Learning-based Automatic ECN Tuning Schemes}
Machine learning, especially RL, provides an effective way to deal with the issues of dynamic schemes.
Researchers have endeavored to explore efficient ways to combine DRL and traditional Active Queue Management (AQM) to deal with network congestion control problems in various dynamic scenarios. For instance, DRL-AQM \cite{ma2021intelligent} adopts a model-free RL method trained in simple network scenarios to capture complex patterns in DCN aiming to improve the performance of DCN.
RL-AQM \cite{alwahab2021deep} combines the deep Q-learning algorithm with the AQM scheme to reduce queuing delays while ensuring good link utilization, targeting to automatically adapt to dynamic network conditions.
However, these solutions are primarily designed to auto-tune packet drop probability settings without involving the ECN threshold setting policies, nor are they specially customized for the data center network.
ACC \cite{yan2021acc}, as the only existing learning-based ECN tuning scheme tailored to the characteristics of the data center network, leverages a MARL algorithm DDQN to adaptively learn the ECN adjustment policies based on real-time network states. Experimental results show that ACC achieves superior performance to both static and dynamic schemes. However, ACC cannot work well in the presence of incast and mixed mice-elephant flows, which are not taken into consideration in its learning algorithm. In addition, ACC uses DDQN as the RL algorithm, which requires global experience replay at tuning intervals, resulting in additional system overhead in terms of memory and bandwidth. 
Comparatively, our PET approach takes these issues into account in learning an optimal tuning policy and uses IPPO as the learning algorithm, which can avoid introducing unnecessary system overhead, with higher learning efficiency.


\section{Preliminaries}
\label{motivation}
This section will provide preliminary knowledge about ECN and incast issues, and describe the design goals of this work.

\subsection{Explicit Congestion Notification}
ECN is recognized as an efficient mechanism to realize a congestion-aware DCN, which involves both the network layer and the transport layer. 
As defined in RFC 3168, in the IP header, the two bits on the right side of the DSCP field are used for ECN, where $00$ indicates a non-ECN-capable-transport (Non-ECT) endhost, $01$ or $10$ indicates an ECN-capable-transport (ECT) endhost, and $11$ indicates congestion encountered (CE). 
Similarly, in the TCP header, the one-bit ECE control field is used to signal an ECN-Echo to notify the TCP sender that the packet has experienced congestion in the network.
The whole working procedure of ECN-based CC schemes can be briefly summarized as follows.
When a sender and receiver are trying to establish a TCP connection, they first need to confirm both sides support ECN through the packets marked with ECT.
At the switch side, each queue of output ports is configured with a pre-defined threshold $K$ (or two thresholds $K_{max}$, $K_{min}$). Then, the switch employs the active queue management (AMQ) mechanism to actively mark the packets with the CE codepoint (i.e., setting the ECN field to $11$) whenever the queue length exceeds the threshold $K$. Then, the receiver host acknowledges each CE-marked packet with the ECN-Echo flag (i.e., ECE) in its TCP header and sends this ACK back to the sender host. Lastly, the sender regulates its congestion window size based on these received congestion notifications.


\subsection{Incast Problem}\label{incast}
TCP incast was first identified and described by Nagle et al. \cite{tang2004panasas} in a distributed storage cluster, where files are intentionally stored in multiple servers. TCP congestion may occur at or near the receiver when multiple blocks of a file fetched from multiple servers are sent back to the sender at the same time, which is known as TCP incast.
The data center network incast problem refers to the fact that, for the partition-aggregation distributed computing paradigm, many flows often happen to be aggregated on the same egress interface of the last hop switch in a short period of time, at which time packets may overflow the switch buffer, resulting in packet loss. This can happen even if the traffic is small. This problem arises naturally from the partition-aggregation traffic pattern, which can significantly degrade the performance of data center networks. There have been many ECN-based schemes to alleviate the incast problem. 
For example, the DCTCP protocol uses the ECN mechanism to mark packets in the switch queue that exceed the marking threshold, and restrains the sender's sending rate according to the proportion of marked packets so as to avoid queue overflow on the switch, thereby alleviating the incast problem. 
However, how to set an appropriate ECN marking threshold remains a challenging task, where the ECN settings can largely affect the performance of the algorithm.
This is also the main problem that PET aims to address.

\subsection{Design Goals} 
In our PET approach, we have three main design goals:

\begin{itemize}
    \item \textbf{Goal 1:} To provide a near-optimal ECN tuning policy based on more comprehensive knowledge about the dynamic network conditions so as to achieve better adaptability to various network dynamics.
    \item \textbf{Goal 2:} To maximize the performance of the used IPPO algorithm to facilitate the ECN tuning, targeting a higher learning efficiency and better robustness.
    \item \textbf{Goal 3:} To minimize the overhead caused by the learning algorithm, such as the memory and bandwidth overhead resulting from the experience replay.
\end{itemize}

\section{PET Approach Design}
\label{approach}
In this section, we elaborate on the design of our PET approach. 
Inspired by ACC \cite{yan2021acc}, which formalizes the ECN marking threshold problem in high-speed DCNs as an RL problem, we design our PET based on the more efficient MARL algorithm IPPO, which can automatically tune the ECN marking threshold without human intervention to configure ECN settings when the network condition changes. 
In addition, compared with ACC, we also extend to consider more contributing factors to congestion, which will be used as the inputs of the multi-agent DRL model, aiming to better characterize the constantly changing network states and generate better ECN tuning policies.
Consistent with ACC, PET is easy to be deployed in production data centers and can run with commodity hardware without requiring modifications to existing network stacks.

%

\subsection{Design Overview}

\subsubsection{\textbf{Reinforcement Learning Modeling}}
\begin{figure}
	\begin{center}
		\includegraphics[width=8cm]{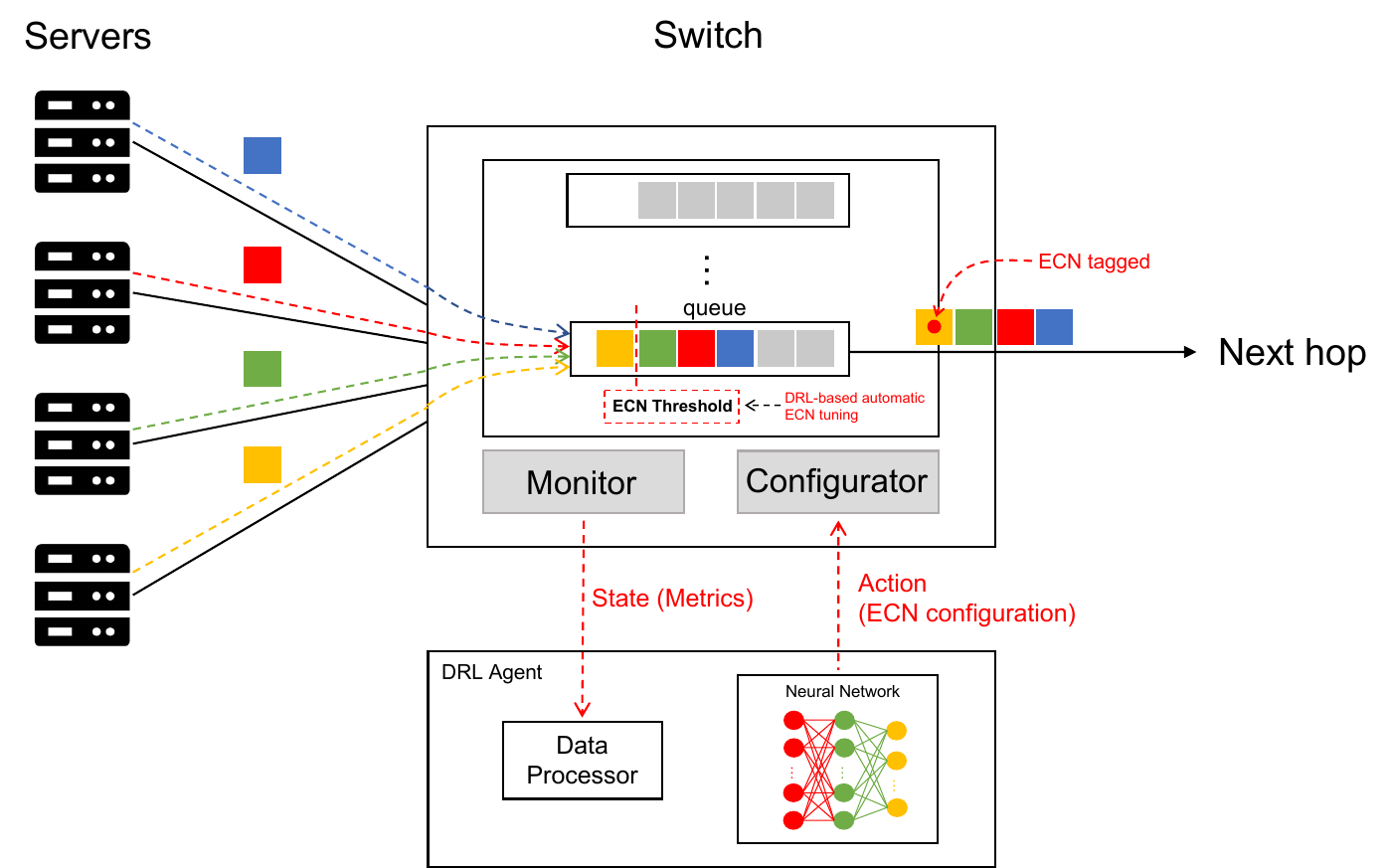}
		\caption{Reinforcement learning modeling}\label{Switch}
	\end{center}
	\vspace{-0.4cm}
\end{figure}

The problem of dynamically tuning ECN marking thresholds can be formulated as an RL problem, as depicted in Fig. \ref{Switch}.
ECN-based schemes have been extensively studied in data centers to alleviate the incast problem.
These schemes mark the packets in the switch queue that exceed the ECN marking threshold and limit the sending rate of the sender based on the proportion of marked packets to prevent overflowing the switch queue. 
However, traditional non-learning-based ECN threshold settings cannot well adapt to network dynamics. 
RL is a learning paradigm that aims to learn an optimal policy by allowing an agent to interact with the environment and take appropriate actions based on current states to maximize the gain or return.
One RL task can be theoretically characterized as a Markov Decision Process (MDP), which defines a mathematical model that can be used for the optimal decision process of stochastic dynamic systems. 
Typically, an MDP can be represented by a 5-tuple, as given below:
\begin{equation}
\label{fivetuple}
\mathcal{M}=<\mathcal{S}, \mathcal{A}, \mathcal{R}, \mathcal{P}, \gamma>,    
\end{equation}
where $\mathcal{S}=\{s_1,s_2,...,s_n\}$ is the state space that denotes the network conditions counted by the agent. $\mathcal{A}=\{a_1,a_2,...,a_n\}$ is a set of actions, which here specifically refer to the ECN marking thresholds outputted by the model. 
In particular, we use discrete ECN marking thresholds to reduce the size of the action space. 
$\mathcal{R}=\{r_1,r_2,...,r_n\}$ denotes the set of rewards, where $r_t$ represents the reward received by the agent after taking action $a_t$ resulting in a transition from state $s_t$ to state $s_{t+1}$.
$\mathcal{P}(s_t,a_t)$ is the transition probability from state $s_t$ to $s_{t+1}$ after taking action $a_{t}$. $\gamma \in[0,1]$ is a discount factor, which controls the extent to which we favor immediate returns over returns from the distant future.
The objective of an RL agent is to select the best action that can receive the highest reward in each state so as to maximize the cumulative reward over the long term.


\subsubsection{\textbf{Implementation Choices}}
In the DCN scenario, each switch acts as an agent of RL. As there are many switches in DCN, thus precisely it is a multi-agent RL problem.
There are usually two ways to implement MARL algorithms, i.e., Centralized Training with Decentralized Execution (CTDE) and Decentralized Training with Decentralized Execution (DTDE).
Intuitively, it is better to directly adopt the CTDE framework, which is centrally trained based on the global state information collected from all switches for the sake of improving the accuracy of the algorithm to the maximum extent. However, online centralized training in real time is impractical in a large-scale DCN due to the following three issues.
\begin{enumerate}[(i)]
    \item \textit{Difficult to Converge.}
    The CTDE framework requires to collect the state information of all switches in the network and distribute the MARL outputted corresponding ECN settings to switches for execution. However, for a large-scale DCN, the huge state space and action space will make the MARL model difficult to converge. 
    \item \textit{Bandwidth Overhead.}
    Given the huge state space and action space, constantly collecting information from all switches and distributing policies from the central server to switches will occupy substantial bandwidth resources, which will negatively impact the transmission of real service data traffic.
    \item \textit{Slow Reaction.}
    The state information collection and the distribution of ECN configurations will take a relatively long time, which is unacceptable in a high-speed DCN. Worse still, due to such a long time of information collection and policy distribution, when these configurations arrive at respective switches, they may no longer apply to the current network states due to the rapidly changing DCN environment. Note that this problem will be exacerbated if there is congestion in the network.
\end{enumerate}

As a result, the DTDE framework is better suited to the dynamic ECN tuning problem in high-speed DCN at runtime. In practice, each switch is associated with a DRL agent, and all these distributed agents form the multi-agent system. Each DRL agent observes the local network state and independently selects the best action to execute based on the reward function. 
Only using local network states will greatly reduce the state and action space, which enables the learning process of individual agents to converge faster. Moreover, the DTDE framework can also eliminate the transmission latency and bandwidth cost by avoiding inter-device communication.
In view of these benefits, in our PET approach, the DTDE framework will be adopted.


\subsection{Problem Formalization}
In the DTDE-based MARL model, each distributed agent equipped on the switch will be trained based on the collected network state information from the local port, and then executes ECN configuration decisions generated by the local learning model.
Specifically, we will use the IPPO as the learning algorithm, which is a type of independent learning algorithm that can perform better than state-of-the-art joint learning methods with little hyperparameter tuning. More details about the IPPO algorithm are provided in Section \ref{algorithm}.
The MDP modeling is built as follows.

\subsubsection{\textbf{States}}
The states represent the information about the network conditions that are used as inputs to the learning agent. 
Note that it is not that the more states the better, because a larger dimensional state space will be detrimental to training efficiency, and if in case some unuseful or even confounding factors are selected, it will negatively affect the quality of trained policy.
Therefore, we set the following five principles for selecting appropriate states:
\begin{enumerate}[(i)]
    \item The selected metrics need to be able to characterize the state of the network. 
    Typically, these metrics are oriented toward network throughput exposing the status of network utilization, which can establish a network statistics baseline for evaluating the network condition.
    The output rate of each link is such a typical representative metric to meet with this guideline.
    \item The selected metrics need to accurately represent the congestion condition in the network. 
    These metrics can aid the algorithm in better comprehending the congestion condition in the network, allowing it to tune the ECN threshold with congestion awareness. The current queue length and the current ECN threshold in the switch queue, which can give immediate feedback on the accurate state of network congestion, 
    are the most appropriate and straight metrics to achieve this goal.
    \item The selected metrics need to characterize the network incast issue, which is critically detrimental to network congestion.
    This type of metrics can enable the algorithm to tune the ECN threshold with incast awareness, thus ultimately enhancing the data center network's ability to be aware of and be able to handle the intractable incast problem.    
    \item The selected metrics need to provide information about traffic diversification so as to enable the ECN tuning algorithm to meet the differentiated service requirements of different types of traffic (i.e., mice flows and elephant flows).
    \item The selected metrics should be simple and compatible with the majority of commodity  switches. The computation of the metrics must be carried out in a way that does not consume switch resources excessively, so as to protect the switch's performance.
\end{enumerate}

Therefore, in the light of the above principles, we select six pivotal factors that contribute most to network congestion to constitute the state space.
These collected statistics will be first pre-processed to make them compatible with switches of different vendors.
Here, we divide the six pivotal factors into two categories. The first category of factors characterizes the basic network conditions that are readily available in the switch, including the current queue length ($qlen$), the output data rate for each link ($txRate$), the output rate of ECN-marked packets for each link ($txRate^{(m)}$) and the current ECN threshold ($ECN^{(c)}$). 
The second category of factors denotes the network characteristics that can be obtained by simple calculations, which includes the incast degree ($D_{incast}$) and the current ratio of mice and elephant flows ($R_{flow}$). 
The above six key factors will help the learning agent to better understand the real-time network dynamics. 
Next, how to quickly obtain the metrics that are not directly available in the switch has become a new major concern. 
For the incast degree, we will analyze the packet headers and determine the sender and receiver based on the packet header information. Then, we will calculate the total number of senders communicating with the same receiver in each many-to-one traffic pattern, and output this number as the incast degree.
For the ratio of mice and elephant flows, for the sake of simplicity, we first apply the easiest way to distinguish them based on the size of flow, where one flow whose cumulative size exceeds 1 MB will be recognized as an elephant flow according to the rule in \cite{curtis2011devoflow}.
Then, the ratio of mice and elephant flows will be calculated accordingly.
Note that one can also use other flow classification methods, which is not the focus of this paper. 
To sum up, we use a six-tuple to represent the network state $s_t$ at the time slot $t$, i.e.
\begin{equation}
s_{t} = (qlen,txRate,txRate^{(m)},ECN^{(c)},
 D_{incast},R_{flow}).
\end{equation}
It makes sense to provide the normalized values to agents rather than directly provide the values of collected statistics because normalization helps agents to generalize to different network environments. 
In addition, to measure the changes in the statistics collected over consecutive time slots, we use the queue states of the last $k$ monitored time slots as the time sequence state information for each tuning inference. 
The sequence state $s'_{t}$ at the time slot $t$ can be represented by:
\begin{equation}
\label{sequence_state}
s'_{t} = \{s_{t-k+1},...,s_{t-1},s_{t} \} \in \mathcal{S}
\end{equation}

\subsubsection{\textbf{Action}}
The actions of the agent are defined as the ECN settings in the switch. 
Here, we apply the ECN parameter settings in AQM scheme, which include a high marking threshold ($K_{max}$), a low marking threshold ($K_{min}$), and a marking probability ($P_{max}$).
\begin{equation}\label{actions}
a_{t} = \{K_{max},K_{min},P_{max} \}
\end{equation}

To reduce the action space, we refer to the practice of ACC to discretize the continuous action space into discrete action space for setting the ECN marking thresholds.
Specifically, we use an 
exponential function $E(n)$ to determine the discrete action values (i.e., $K_{min}$, $K_{max}$):
\begin{equation}\label{exponetialfunction}
E(n)=\alpha \times 2^{n} KB,
\end{equation}
where $\alpha$ is a scale parameter. The calculated $K_{min}$ is ensured to be less than $K_{max}$. 
Considering the computing complexity, it is recommended to limit the value of $n$ to a small value, e.g., $n\in[0,9]$.
As for the discretized marking probability $P_{max}$, in our simulations, we set its tuning interval to 5\%.


Besides, too frequent ECN marking threshold tuning operations can impose high pressure on the switch and cause performance oscillations, thus we set a time parameter $\Delta t$ to restrict the time interval between two adjacent tuning operations. 
A typical value of $\Delta t$ is an order of magnitude greater than the RTT, which can effectively avoid causing negative impacts on switch performance.


\subsubsection{\textbf{Reward}}
The reward function provides an incentive mechanism using reward and punishment to guide agents in learning an optimal policy.
As our PET is implemented based on the DTDE framework, working in a distributed manner. Thus, it is difficult for an agent to obtain the end-to-end latency and overall throughput to be used as the reward since it only has a local observation.
To this end, in the reward function design of PET, we use the link utilization and queue length that can be locally observed to characterize the throughput and packet delay, respectively.
The reward function is defined as:
\begin{equation}\label{rewardfunction}
r = \beta_{1} \times T + \beta_{2} \times L_{a},
\end{equation}
\begin{equation}
T = \frac{txRate}{BW},
\end{equation}
\begin{equation}
L_{a} = \frac{1}{ queueLength_{avg}},
\end{equation}
where $T$ indicates the link utilization and
$L_{a}$ denotes the reciprocal of the average queue length that is used to quantify the delay. $\beta_{1}$ and $\beta_{2}$ are two weighting parameters that are used to balance the tradeoff between throughput and delay, where $\beta_{1}+\beta_{2}=1$.
The weighting parameters can be set according to different traffic requirements to accommodate the differentiated application demands, e.g., a data center with Web Search applications can set a larger $\beta_{2}$ to guarantee a low latency while storage systems can configure a larger $\beta_{1}$ to ensure high throughput.


\subsection{Multi-agent DRL Algorithm}\label{algorithm}
\subsubsection{\textbf{Learning Algorithm Choice}}

Considering that PET uses discrete action space, the intuitive idea is to apply a Deep Q-Network (DQN)-like algorithm, such as the DDQN as used in ACC, since its features well fit the discrete action space and performs better in such a scenario. 
Besides, since each switch acts as an agent, thus the data center network is actually a multi-agent scenario. In this multi-agent scenario, the DDQN-based ACC scheme achieves good performance by relying on global experience replay, which requires an additional replay memory in addition to local replay memory. However, this inevitably introduces memory overhead and extra bandwidth cost in exchanging the experience replay between switches.
Based on this observation, we intend to avoid using the global experience replay schemes, instead, we turn our attention to the state-of-the-art PPO \cite{schulman2017proximal} algorithm for tuning parameters.
The recently proposed multi-agent PPO (MAPPO) \cite{yu2021surprising} has been proved to be an efficient multi-agent variant of PPO. 
However, the MAPPO algorithm is implemented based on centralized training, where multiple individual PPO agents are enforced to cooperate with each other through a global value function.
This violates the distributed design principle of our PET approach, which is implemented based on the DTDE framework. 
Ultimately, we choose the decentralized IPPO\cite{de2020independent} as our learning algorithm, which proves to be robust to the nonstationarity of the environment and outperforms state-of-the-art multi-agent RL CTDE schemes.




In essence, multi-agent IPPO is a kind of independent learning algorithm, in which each distributed agent (switch) learns independently based on its local observation history 
and estimates its local value function, without the need of global experience replay.
In our PET framework, each switch executes the IPPO algorithm independently and learns a critic, which can be expressed as the function $V_\omega(s_t)$ parameterized by $\omega$ utilizing Generalized Advantage Estimation (GAE) \cite{schulman2015high}. Moreover, each switch has an advantage estimation function $\hat{A_t}$ with discount factor $\gamma$ and a hyperparameter $\lambda$,
which is defined  as follows:
\begin{align}
\hat{A_t} = \delta_t + (\gamma\lambda)\delta_{t+1} + \cdots  + ({\gamma\lambda})^{T-t-1}\delta_{T-1},
\end{align}
where
\begin{align}
\delta_t &= r_t + \gamma{V_\omega(s_{t+1})} - V_\omega(s_t),
\end{align}
To learn a policy $\pi$ for switches, the policy loss is given as:
\begin{align}
\label{policy_loss}
L_t^\pi(\theta)& =  \mathbb{E}_t [min(\frac{\pi_\theta(a_t|s_t)}{\pi_{\theta_{old}}(a_t|s_t)}\hat{A}^{\pi_{\theta^{}}}(s_t,a_t), clip(\frac{\pi_\theta(a_t|s_t)}{\pi_{\theta_{old}}(a_t|s_t)}, \\ \nonumber
&1-\epsilon, 1+\epsilon)\hat{A}^{\pi_{\theta^{}}}(s_t,a_t))],
\end{align}
where $\pi_{\theta_{old}}$ represents a policy parameterized by $\theta_{old}$, $\theta$ represents the policy parameters.
The value estimation needs to minimize a square-error loss, as given below:
\begin{align}
\label{value_loss}
L_t^v(\omega) = \mathbb{E}_t [ (V_\omega(s_t) - \hat{R_t})^2],
\end{align}
where $\hat{R_t}$ is the sum of the rewards obtained from the environment from time $t$. 
Algorithm \ref{alg:PPO} describes the multi-agent IPPO-based learning algorithm used in our PET. 

\begin{figure*}
	\centering
		\includegraphics[width=0.75\linewidth]{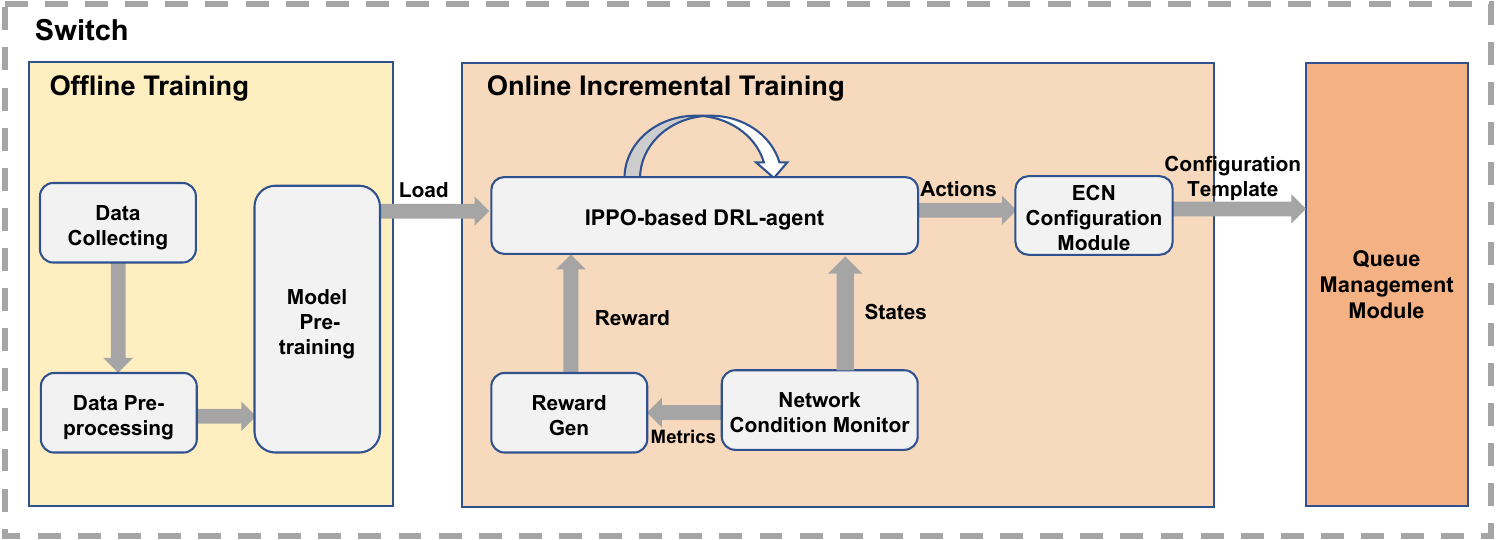}
		\caption{Overview of PET architecture}
		\label{Architecture}
		\vspace{-0.5cm}
\end{figure*}
\begin{figure}[!t]
    \renewcommand{\algorithmicrequire}{\textbf{Input:}}
	\removelatexerror
	\begin{algorithm}[H]
	\caption{PET's Learning Algorithm based on Multi-agent IPPO}
        \label{alg:PPO}
        \LinesNumbered
		\KwIn{initial policy parameters $\theta_{0}$, initial value function parameters $\psi_{0}$}  
		\For{episode = 1,2,...}
		{
		Reset the environment with the initial state,$s_{init}$\;
		Sample the dynamics parameters $\lambda$ from a range $\gamma$ uniformly, $\lambda \sim \gamma$\;
    		\For {switch i= 1,2,...}{
    		Collect set of trajectories  $\mathcal{D}_{k}=\left\{\tau_{i}\right\}$  by running policy  $\pi_{k}=\pi\left(\theta_{k}\right)$  in the environment\;
    		Compute rewards-to-go  $\hat{R}_{t}$ \;
    		Compute advantage estimates,  $\hat{A}_{t}$ based on the current value function $V_{\phi_{k}}$\;
             Optimize policy loss $L_t^\pi(\theta)$ via Equation \ref{policy_loss} with Adam optimizer and learning rate  $l_{a}$  for  $N$  epochs\;
    		$\theta_{\text {old }} \leftarrow \theta $\;
    		
            
            Optimize policy loss $L_t^\pi(\omega)$ via Equation \ref{value_loss} with Adam optimizer and learning rate  $l_{v}$  for  $N$  epochs\;
    		$\psi_{\text {old }} \leftarrow \psi $\;}
    	}
	\end{algorithm}
 \vspace{-0.3cm}
\end{figure}

\subsection{Hybrid Training of Multi-agent DRL Algorithm}
One problem with deep reinforcement learning is the need for risky trial-and-error during the online learning process, which may result in irreversible effects and severely impact the performance of data center networks. 
For this reason, we adopt a hybrid training strategy that combines offline training and online training to adaptively accommodate network dynamics and make the model more efficient.
When deploying PET in real-world data centers, it is expected to firstly offline pre-train the model based on an appropriate amount of collected historical network statistics to obtain an initial model.
After offline training, the pre-trained initial model will be installed on switches, which then use local network state information to incrementally train its own local model online to improve the overall quality of the model. 
In the early stages of online training, the pretrained model requires a high exploration rate to rapidly adjust its parameters. However, as training progresses, the model gradually stabilizes. At this point, we aim to maintain a lower exploration rate to ensure the stability of the model's output strategy. This represents a delicate balance between exploration and exploitation. Consequently, the probability (i.e., $\epsilon$ defined in Eq.(\ref{policy_loss})) of selecting exploration actions decays exponentially and actions that yield large rewards are prioritized during online training. 
We use Eq. \ref{epsilon_decay} to exponentially decay the exploration rate. 
\begin{align}
\label{epsilon_decay}
\epsilon_{t}=decay\_rate ^{\frac{t}{T}} *  \epsilon \ \text{when}\ t>T,
\end{align}
where $\epsilon_{t}$ represents the value of $\epsilon$ at the $t$-th  training step, $t$ is the count of training steps starting from 0, and T represents the decay step used to control the descent rate of the $\epsilon_{t}$.
 Fig. \ref{Architecture} depicts the general architecture of our PET approach, which implements automatic ECN tuning based on multi-agent IPPO. The whole architecture mainly consists of two parts, i.e., offline training and online incremental training.


\subsubsection{Offline Training}
In the offline training part, we should first collect the training data, which are either historical network state information collected from the switches deployed in the current data center or simulated traffic data conforming to the data distribution of the existing network. Next, the collected data will be pre-processed and then used for model pre-training.
Overall, it is reasonable and necessary to perform offline model pre-training since directly online training the model from scratch in a real DCN environment will inevitably lead to excessive trial and error costs, which will eventually seriously affect the effectiveness of the learning algorithm.
Note that the pre-trained initial model cannot be fully adapted to the specific scenarios of all switches. Subsequently, the pre-trained initial model will be deployed on each switch for further online local incremental learning.

\subsubsection{Online Incremental Training}
In the online training part, the pre-trained initial model will be pertinently enhanced by local training to suit the specific environment of each switch. 
Specifically, the IPPO-based DRL agent performs incremental local training by interacting directly with the environment to optimize actions based on local network state statistics, which are collected by the Network Condition Monitor (NCM). 
Besides, NCM is also responsible for handling the delay and bandwidth information that will be transferred to the Reward Generation module.
The DRL agent then outputs the corresponding discretized actions (i.e., values of $n$ defined in Eq.(\ref{exponetialfunction}) and $P_{max}$ defined in Eq.(\ref{actions})) to the ECN Configuration Module (ECN-CM). Afterwards, ECN-CM calculates the ECN marking thresholds (i.e., $K_{min}$ and $K_{max}$ defined in Eq.(\ref{actions})) based on the received actions, and finally generates an ECN configuration template that will be delivered to the Queue Management Module (QMM) to perform ECN threshold configurations on queues.

{\subsection{Implementation in Real-world Scenarios}When implementing in real-world scenarios, it is crucial to prioritize the specific requirements and address infrastructure cost concerns. 
This subsection will detail the implementation of the Network Condition Monitor module, which is equipped on each device to facilitate requisite calculations and alleviate the memory cost of switches. Additionally, we will discuss how to adapt our PET in multi-queue scenarios.

{\subsubsection{Network Condition Monitor}
As shown in Fig. \ref{Architecture}, the Network Condition Monitor (NCM) serves as a vital module in online incremental training and plays a crucial role in enhancing its performance. In general, NCM performs the following three main roles:
\begin{itemize}
\item[1)] Monitoring: The NCM periodically retrieves status information from the queues, including network condition information inputted to the DRL agent and queue metrics information fed to the reward generation module. After collection, this information undergoes simple processing by the computation and analysis module. Finally, expired information is promptly cleaned by the scheduled cleanup module to minimize memory costs.
\item[2)] Computation and Analysis: The computation and analysis module performs simple processing on the collected network condition information, primarily focusing on the network condition information provided to the DRL agent. This involves performing calculations to determine the extent of incast and the ratio of elephant flows to mice flows. After normalizing these pieces of information, they are passed on to the DRL agent to generate appropriate execution policies.
\item[3)] Scheduled Cleanup: The scheduled cleanup module is responsible for removing expired network condition information from the monitoring module.
As defined in Eq.(\ref{sequence_state}), the network condition information dating back to slot $t-k$ and earlier can be regarded as expired at slot $t$. Likewise, for the queue metrics information fed to the reward generation module, data before slot $t$ are considered invalid. To ensure the timely removal of expired information, we utilize two cleanup strategies. The first strategy is the scheduled cleanup strategy, which initiates periodic cleanup tasks at regular intervals to remove expired information and free up switch memory. The scheduled cleanup strategy is particularly effective in maintaining a relatively low level of switch memory utilization during normal traffic conditions. However, in scenarios involving incast, where there are sudden bursts of traffic, the scheduled cleanup strategy alone may not be sufficient. To address such cases, we activate the threshold cleanup strategy. This strategy is triggered when the memory utilization reaches a certain threshold.  It removes a proportionate portion of the expired information, thereby freeing up switch memory. With the assistance of these two strategies, we ensure that the switch does not consume excessive memory even during bursts of traffic.
\end{itemize}
Benefiting from the NCM, the online incremental training process can efficiently acquire network states and metrics while mitigating memory consumption concerns. Moreover, when transitioning from a single-queue to a multi-queue configuration, it is easy and straightforward to adapt the Monitoring component of NCM to extract the desired information from the multiple queues.

{\subsubsection{Adaption to Multi-queue Scenarios}
PET was initially designed as a single-queue approach, but it can be readily adapted to accommodate a multi-queue scenario through straightforward adjustments. 
The key modifications involve adjusting the input and output of the algorithm.
To support multiple queues, the algorithm needs to incorporate information from all queues by constructing a matrix representation and feeding it as input to the DRL model. 
This adjustment can be made to the monitoring component within the NCM module mentioned above. 
Specifically, during each interval, the NCM module collects network condition information from the multiple queues and transforms it into a matrix format. This matrix is then passed to the DRL agent for further processing. 
Through appropriate computations, the model can generate the output information matrix specific to each queue. Simultaneously, the network condition monitor should aggregate information from all queues to provide input to the reward generator, which can also be achieved by adjusting the monitoring component within the NCM module. 
With these simple adjustments, the DRL model can be adapted and adjusted based on the obtained reward values. 
These simple adjustments enable a seamless transition from a single-queue scenario to a multi-queue scenario, without requiring any hardware modifications to the switch configuration.

Moreover, note that our PET primarily focuses on the automatic tuning of the queue ECN thresholds at each switch, without any modifications to queue scheduling or sender-side congestion window adjustments, and it is also not involved in resource allocations, ensuring the fairness of our algorithm towards various protocol traffics during its execution.




\begin{figure}
	\begin{center}
		\includegraphics[width=5.5cm]{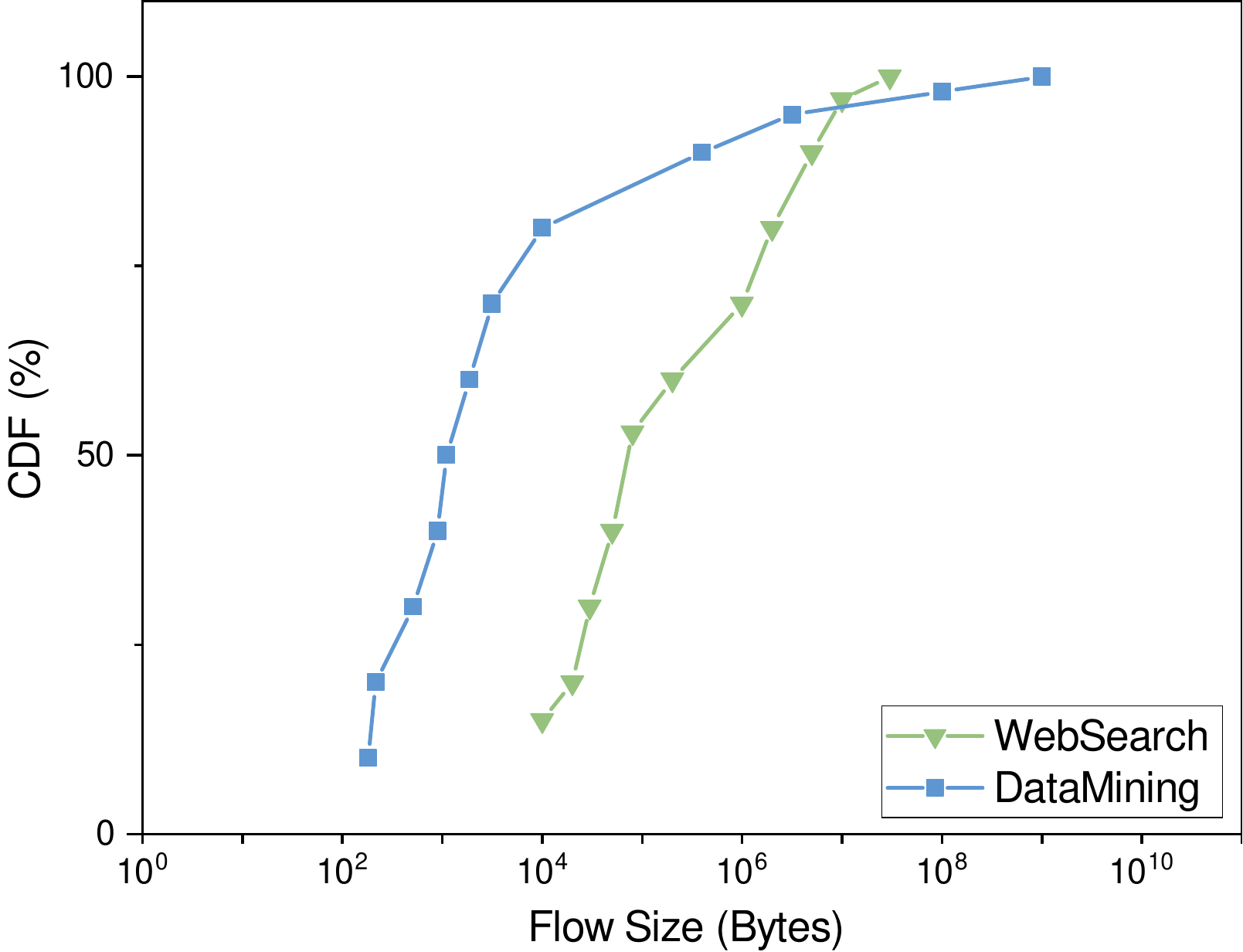}
		\caption{Traffic distributions}\label{traffic}
	\end{center}
	\vspace{-0.5cm}
\end{figure}

\section{Performance Evaluation}
\label{evaluation}
To validate the effectiveness of our PET approach, we implemented a prototype of PET using the ns-3 \cite{ns3} simulator. We conducted extensive simulations to evaluate the performance of PET in a large-scale high-speed RDMA-based data center network and compared it with the state-of-the-art learning-based ACC scheme \cite{yan2021acc} and two representative static ECN schemes, i.e., DCQCN \cite{zhu2015congestion} 
.

\subsection{Experimental Setup}
The experiments are conducted on top of a computer with Ubuntu 20.04 OS, 16GB of RAM, an 8-core i7-9700k CPU, and an Nvidia 2060S GPU. 
The PET prototype is implemented using Python 3.7 based on the Anaconda platform \cite{Anaconda}. The IPPO-based DRL algorithm is developed using the PyTorch package \cite{PyTorch}. 
In order to integrate the RL agent developed by the PyTorch package with the data center network environment built by ns-3, we utilized the ns3-gym \cite{ns3-gym}, which is a framework that integrates both OpenAI Gym and ns-3.


\subsection{Parameter Settings}
To facilitate the comparisons with ACC, we used the same network topology setup as in the ACC's large-scale simulation. Specifically, the leaf-spine topology is used to interconnect 288 hosts with 6 spine switches and 12 leaf switches.
Each leaf switch connects to 24 servers using its 24 25Gbps ports and connects to 6 spine switches using 6 100Gbps ports. The scale parameter $\alpha$ of the exponential function (Eq. (\ref{exponetialfunction})) is set to 20. The weights of the reward function (Eq. (\ref{rewardfunction})) are set to $\beta_1=0.3$ and $\beta_2=0.7$ for Web Search applications and $\beta_1=0.7$ and $\beta_2=0.3$ for Data Mining applications. Similarly, following ACC's recommended values, ACC's weights of the reward function are set to $\omega_1 = 0.3$ and $\omega_2 = 0.7$ for the Web Search applications and $\omega_1=0.7$ and $\omega_2=0.3$ for Data Mining applications. The learning rates of the actor-network and critic-network of the PPO model are $0.0004$ and $0.001$, respectively. The coefficient of GAE adjusted variance and bias in the IPPO model is set to $0.01$.
The $decay\_rate$ in Eq. \ref{epsilon_decay} is set to 0.99, $T$ is set to 50, and clip parameter $\epsilon$ is set to 0.2.

\begin{figure*}[htbp]
	\centering
	\subfigure[Overall]{
		\begin{minipage}[t]{0.23\linewidth}
			\centering
			\includegraphics[width=4.5cm]{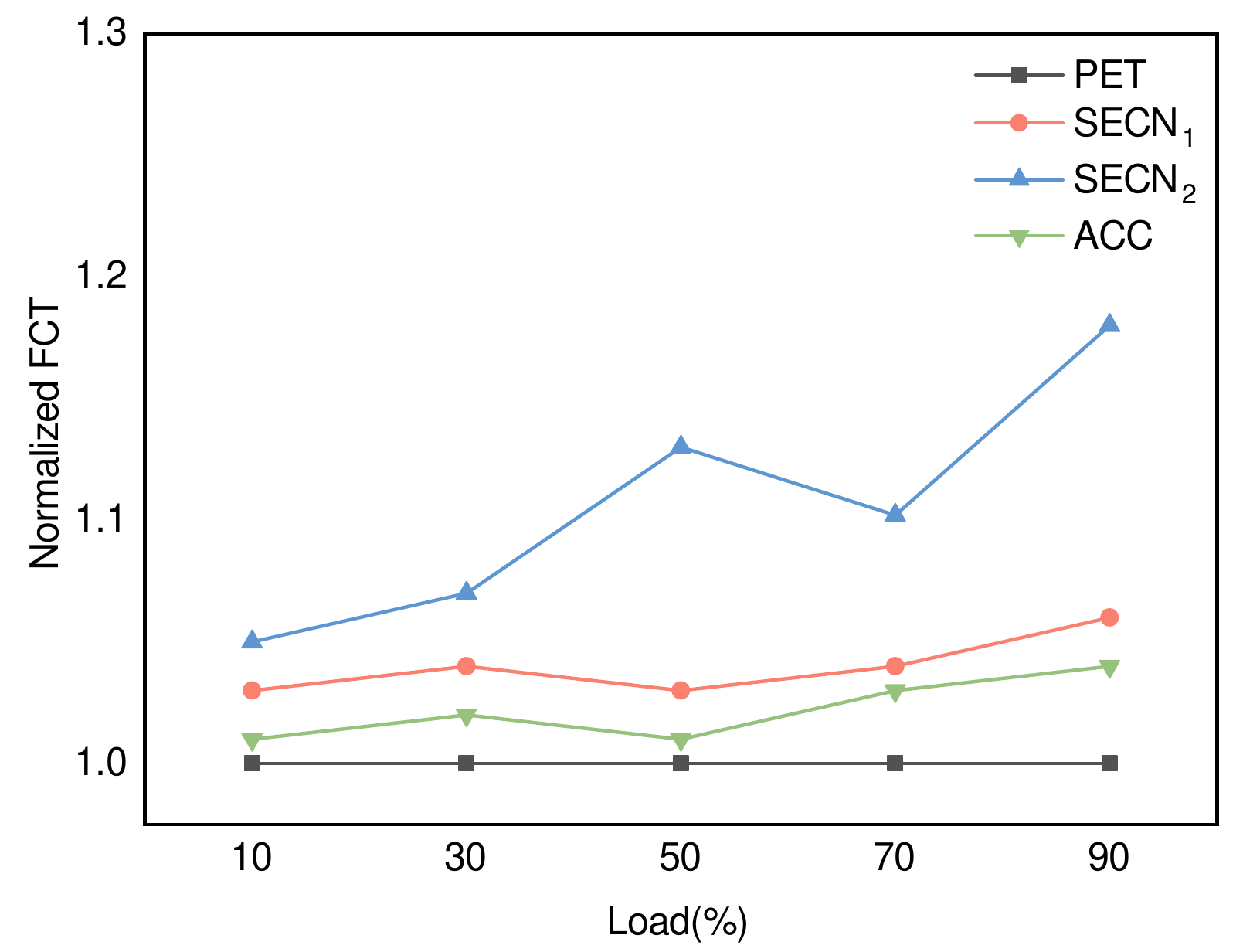}\label{overall}
		\end{minipage}%
	}
	\hspace{0.5mm}
	\subfigure[\text{(0,100KB]:Avg}]{
		\begin{minipage}[t]{0.23\linewidth}
			\centering
			\includegraphics[width=4.5cm]{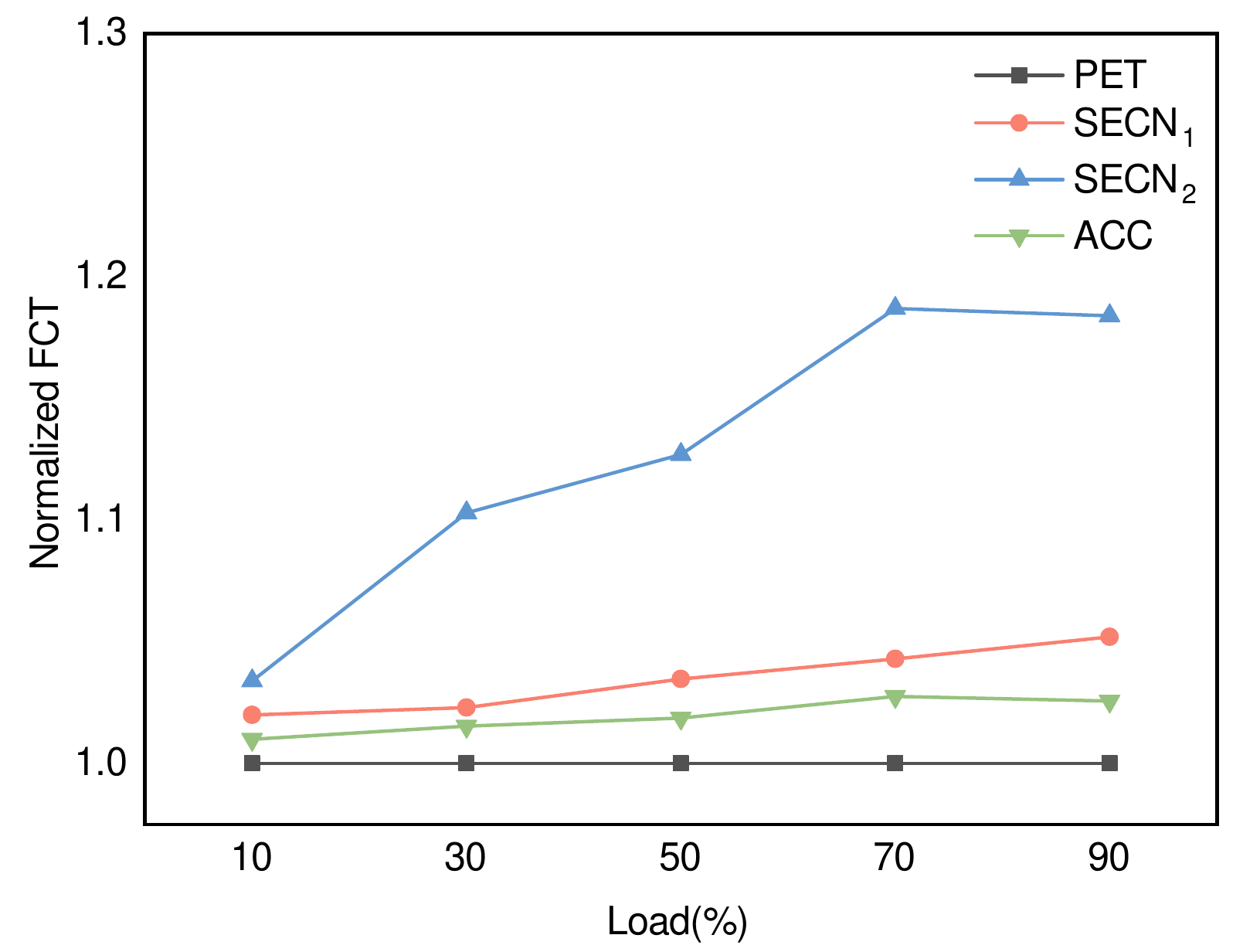}\label{miceavg}
		\end{minipage}
	}
	\hspace{0.5mm}
	\subfigure[\text{(0,100KB]:99th}]{
		\begin{minipage}[t]{0.23\linewidth}
			\centering
			\includegraphics[width=4.5cm]{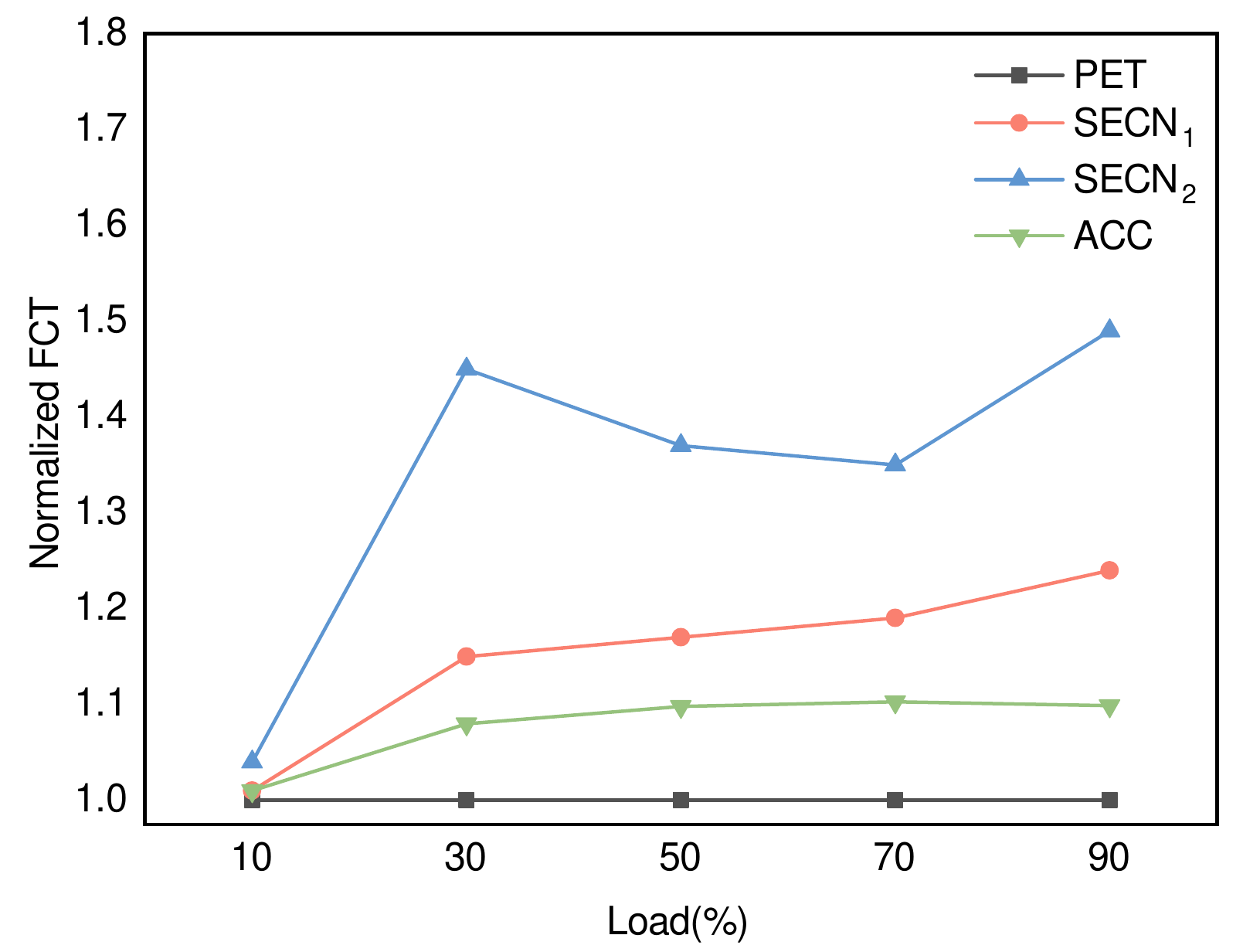}\label{mice99th}
		\end{minipage}
	}
	\hspace{0.5mm}
	\subfigure[[10MB,	$\infty$):Avg]{
		\begin{minipage}[t]{0.23\linewidth}
			\centering
			\includegraphics[width=4.5cm]{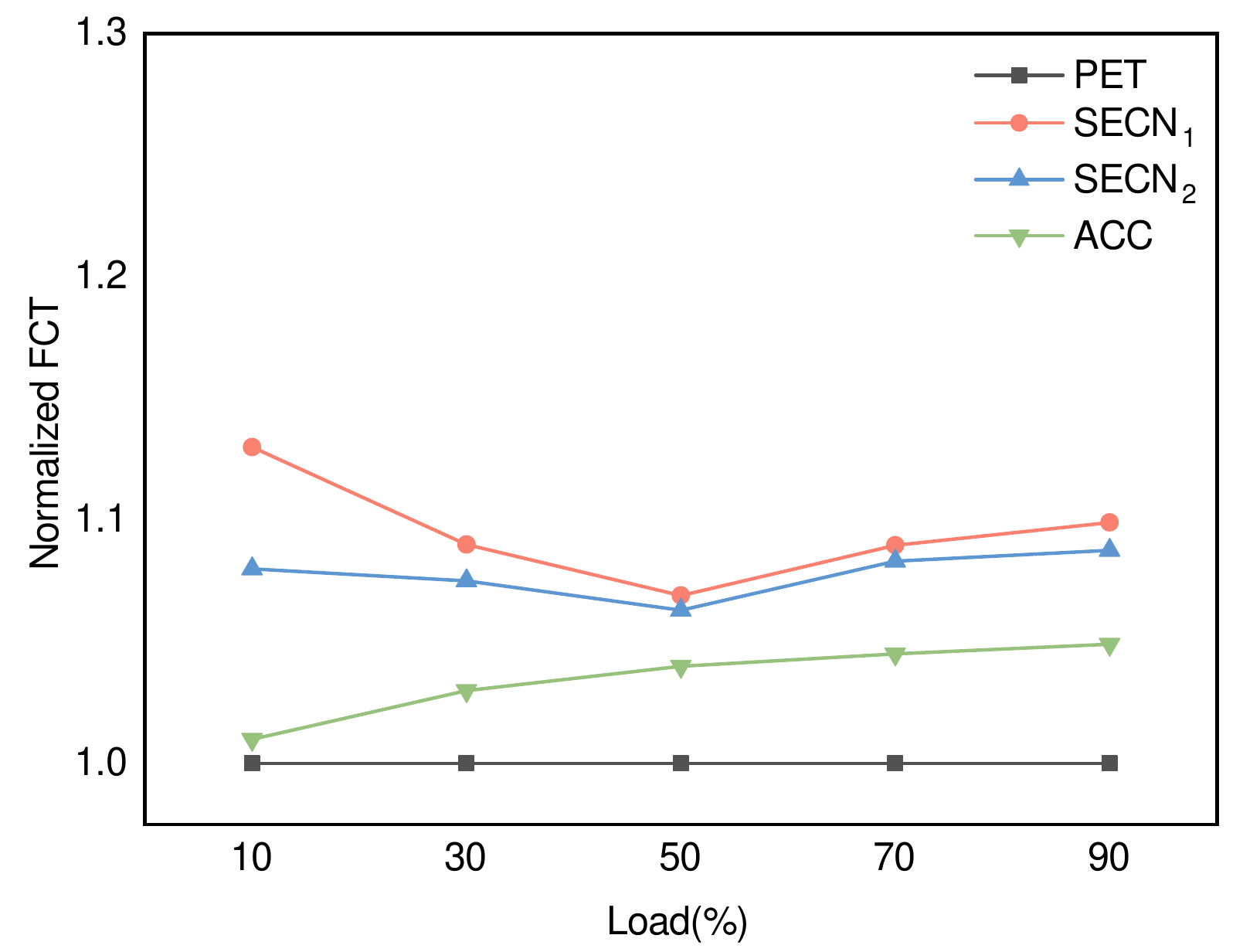}\label{elephantavg}
		\end{minipage}
	}
	\caption{FCT statistics with Web Search workload}
	\vspace{-0.4cm}
\end{figure*}



\subsection{Traffic Generation}
We used the Traffic Generator tool \cite{trafficgen} that is developed by Alibaba for traffic generation. 
Our simulations mainly involve two kinds of datasets, where one is used for offline model pre-training and the other one is used for online incremental training.
For the offline pre-training, we generated the workloads based on two realistic data center application scenarios, i.e., Web Search \cite{alizadeh2010data} and Data Mining \cite{greenberg2009vl2} as shown in Fig. \ref{traffic}, using the traffic generation tool. Note that we extended the Traffic Generator tool to support generating many-to-one traffic patterns and mice and elephant flows. For the online training, similarly, two kinds of workloads, i.e., Web Search and Data Mining, are generated in real-time by invoking the Traffic Generator tool during the run of ns-3.




\subsection{Compared Schemes}
We compare PET with the following three baseline algorithms, i.e., two static ECN schemes and one learning-based ECN tuning scheme.
\begin{itemize}
\item \textbf{Static ECN configuration 1 ($\text{SECN}_{1}$):}  $\text{SECN}_{1}$ refers to the DCQCN algorithm \cite{zhu2015congestion}, where $K_{min}$ is set to $5KB$, and $K_{max}$ is set to $200KB$. 

\item \textbf{Static ECN configuration 2 ($\text{SECN}_{2}$):}  $\text{SECN}_{2}$ refers to the HPCC algorithm \cite{li2019hpcc}, where $K_{min}$ is set to $100KB$, and $K_{max}$ is set to $400KB$. 

\item \textbf{ACC:} ACC \cite{yan2021acc} is the current state-of-the-art learning-based automatic ECN threshold tuning mechanism. 
\end{itemize}


\subsection{Experimental Results}

\subsubsection{Normalized FCT with Web Search Workload}

Fig. \ref{overall} presents the comparison results in terms of FCT with the Web Search workload under different network loads. It can be observed that PET achieves the lowest normalized FCT for all cases. Specifically, taking both mice and elephant flows into account, PET reduces the overall average FCT by up to 3.9\%, 5.8\% and 17.6\% compared with ACC, $\text{SECN}_{1}$ and $\text{SECN}_{2}$, respectively. 
The results in Fig. \ref{miceavg} and \ref{mice99th} 
show that PET can better guarantee the low latency for mice flows than other schemes, where PET achieves up to 2.6\%,  5.2\%, and 18.4\% reduction in the average FCT, and reduces up to 9.9\%, 23.6\% and 48.6\% of 99th FCT for mice flows, compared with ACC, $\text{SECN}_{1}$ and $\text{SECN}_{2}$, respectively.
Fig. \ref{elephantavg} shows the results of average FCT for elephant flows, where PET achieves 4.9\%,  9.6\%, 8.7\% lower average FCT than ACC, $\text{SECN}_{1}$ and $\text{SECN}_{2}$, respectively.

\begin{figure}[htbp]
	\centering
	\subfigure[WebSearch]{
		\begin{minipage}[t]{0.45\linewidth}
		\centering
		\includegraphics[width=4.3cm]{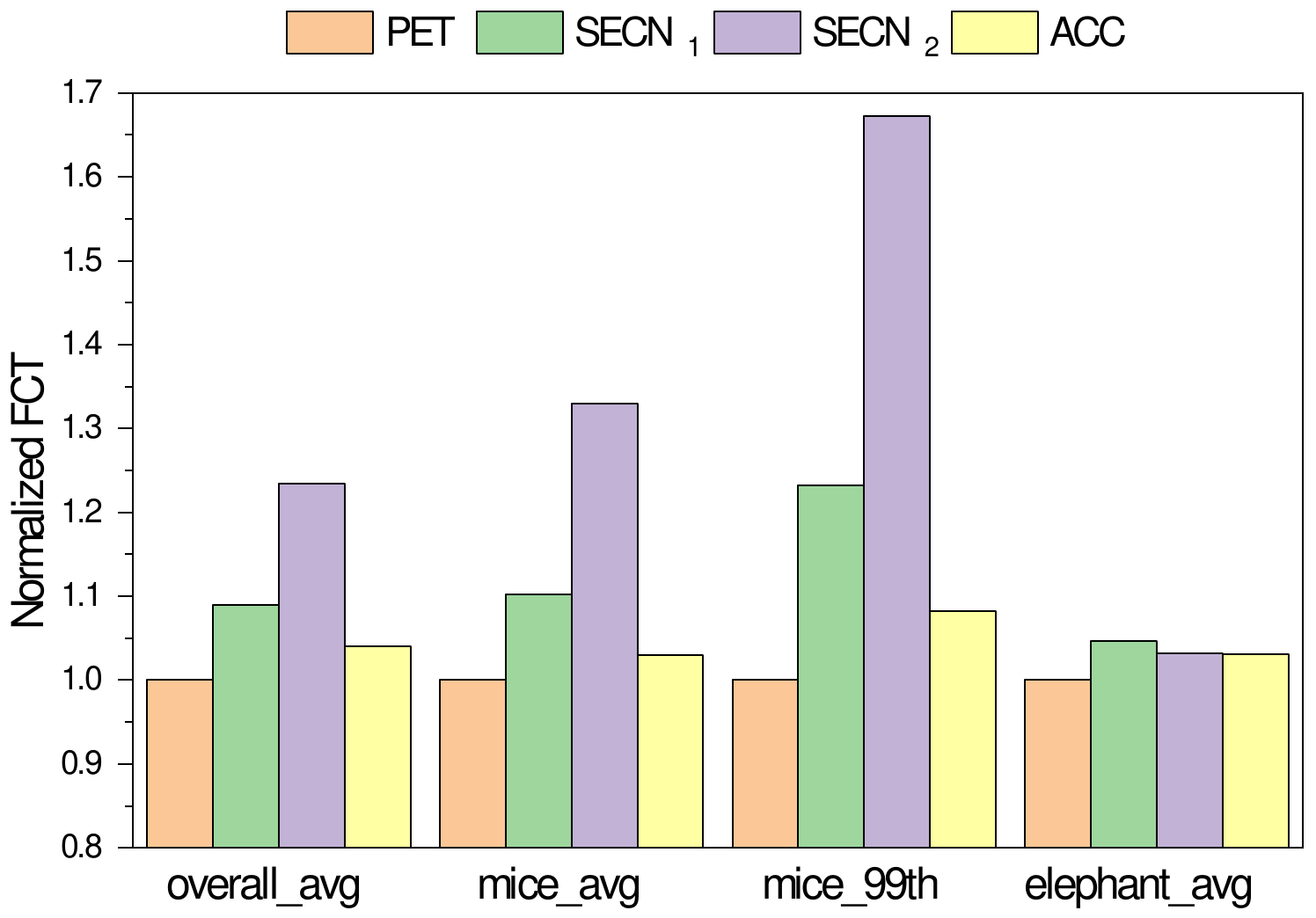}\label{websearch}
		\end{minipage}%

	}
	\hspace{0.5mm}
	\subfigure[DataMining]{
		\begin{minipage}[t]{0.45\linewidth}
			\centering
			\includegraphics[width=4.3cm]{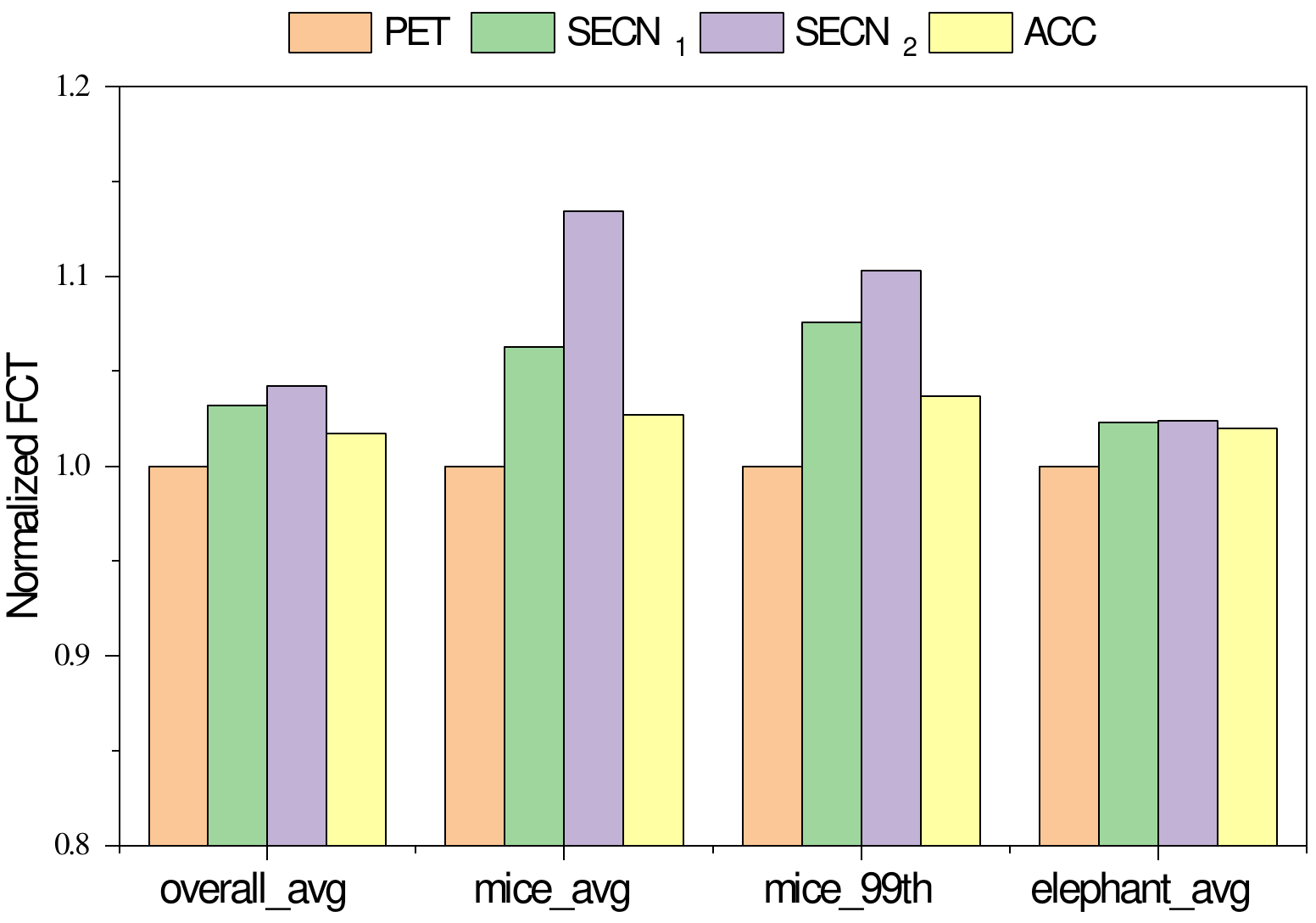}\label{datamining}
		\end{minipage}
	}
	\caption{FCT statistics under different workloads}
    \vspace{-0.5cm}
\end{figure}

\subsubsection{Normalized FCT with Hybrid Different Workloads}
The performance of the model in different scenarios demonstrates its generalization ability to cope with different traffic patterns. 
Thus, we evaluate the performance of PET in two scenarios with different workloads, i.e., Web Search and Data Mining. 
From Fig. \ref{websearch}, we can see that PET achieves the best performance in terms of FCT for all cases in the Web Search scenario, where PET can reduce the FCT by up to  8.2\%, 23.2\% and 67.3\% compared with ACC, $\text{SECN}_{1}$, $\text{SECN}_{2}$, respectively.
This is mainly because more metrics contributing to latency, such as the incast degree, are considered in our PET, which greatly helps the model to better cope with the common incast problem in DCNs thus guaranteeing a low latency. 
Similarly, as shown in Fig. \ref{datamining}, in the Data Mining scenario, PET also achieves the best results. Specifically, PET outperforms ACC, $\text{SECN}_{1}$ and $\text{SECN}_{2}$ with an up to 3.7\%, 7.6\% and 13.4\% reduction in FCT, respectively.
The results prove that PET can achieve a lower latency while guaranteeing high throughput for long flows. The above experiments verify that PET can well cope with different traffic patterns and achieve better performance, which convinces the strong generalization ability of our PET scheme.

\subsubsection{Queue length}
Queue length can be used to verify the PET's performance in terms of latency. If the queue length of the switch keeps at a low level, it implies a better tolerance for incast or bursty traffic, thus ensuring a packet lossless DCN with lower latency.
To this end, we conducted a group of experiments to evaluate the queue length. Table \ref{queuelen} shows the results, from which we can observe that the average queue length for PET is $5.3$KB with a variance of $10.2$KB at 60\% load, while the average queue length for ACC is $6.1$KB with a variance of $14.1$KB. Both PET and ACC can maintain the switch queue at a low level. Nevertheless, the variance of PET is smaller, which indicates that PET is more stable.


\begin{figure}[htbp]
	\centering
	\subfigure[\text{[10MB,$\infty$]:Avg}]{
		\begin{minipage}[t]{0.46\linewidth}
		\centering
		\includegraphics[width=4.2cm]{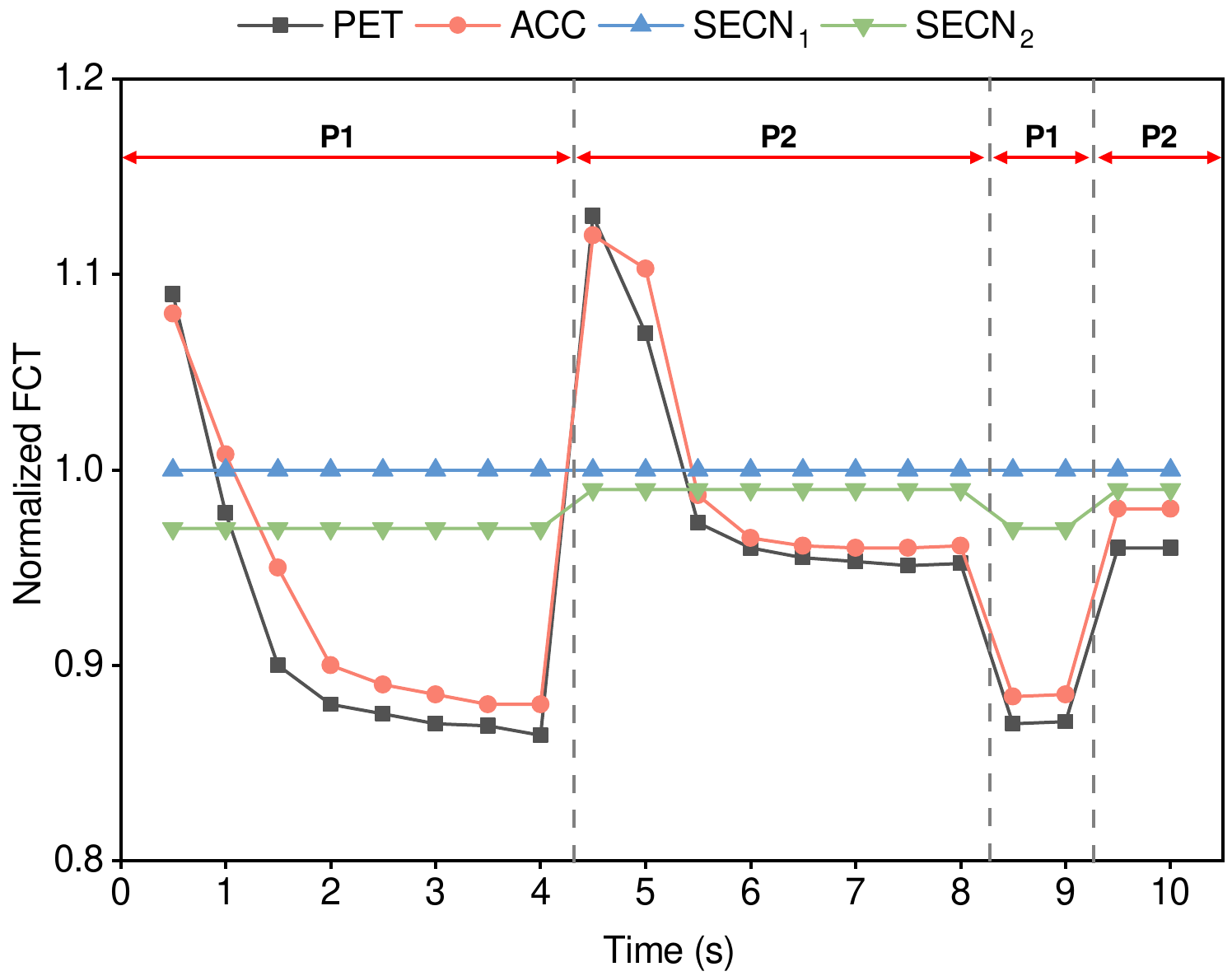}\label{elephantswitch}
		\end{minipage}%
	}
	\subfigure[\text{(0,100KB]:Avg}]{
		\begin{minipage}[t]{0.46\linewidth}
			\centering
			\includegraphics[width=4.2cm]{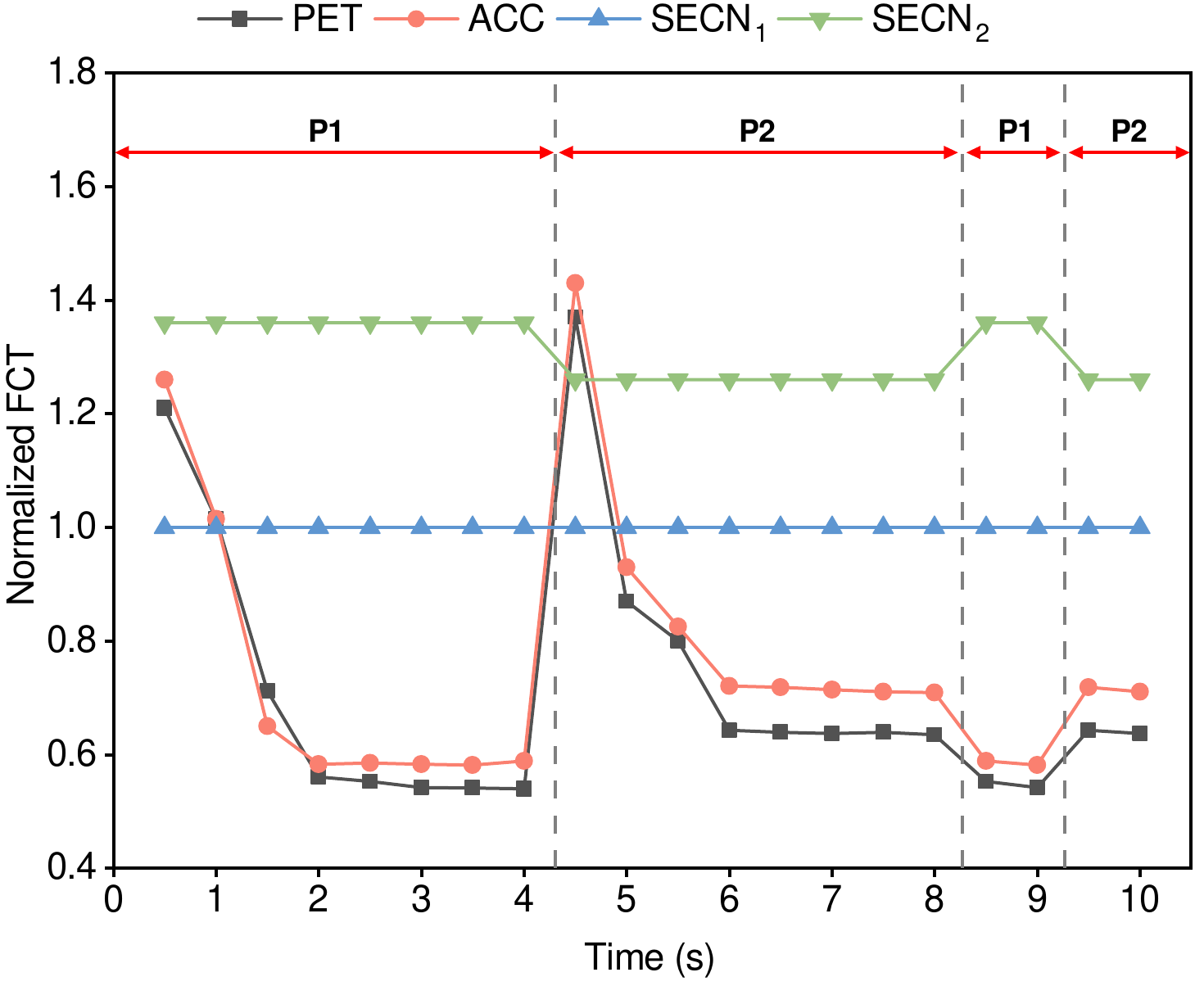}\label{miceswitch}
		\end{minipage}
	}
	\caption{FCT statistics with traffic pattern switching}
	\vspace{-0.3cm}
\end{figure}

\subsubsection{Model Convergence}

The convergence speed of the learning model under changing patterns indicates its ability to adapt to network dynamics and nonstationarity.
To validate the performance of PET in terms of convergence speed, we conducted a series of experiments by abruptly changing the background traffic patterns to see whether PET can quickly converge.
As shown in Fig. \ref{elephantswitch} and Fig. \ref{miceswitch}, we start with the Web Search traffic pattern, and suddenly change the background traffic to be Data Mining traffic pattern at the time 4.1s. At the time 8.1s, we abruptly change the background traffic back to the Web Search traffic, and another sudden traffic switching happens at the time 9.1s. From the results, we can observe that, compared with ACC, PET achieves a 2.1\% and 7.2\% reduction in average FCT for elephant and mice flows in the best case, respectively. The results also imply that both PET and ACC exhibit strong adaptation capability, but comparatively PET achieves lower FCT after fast adaptation, indicating that the ECN setting policies generated by PET are more reasonable than that of ACC.


\begin{figure}
	\begin{center}
		\includegraphics[width=5.5cm]{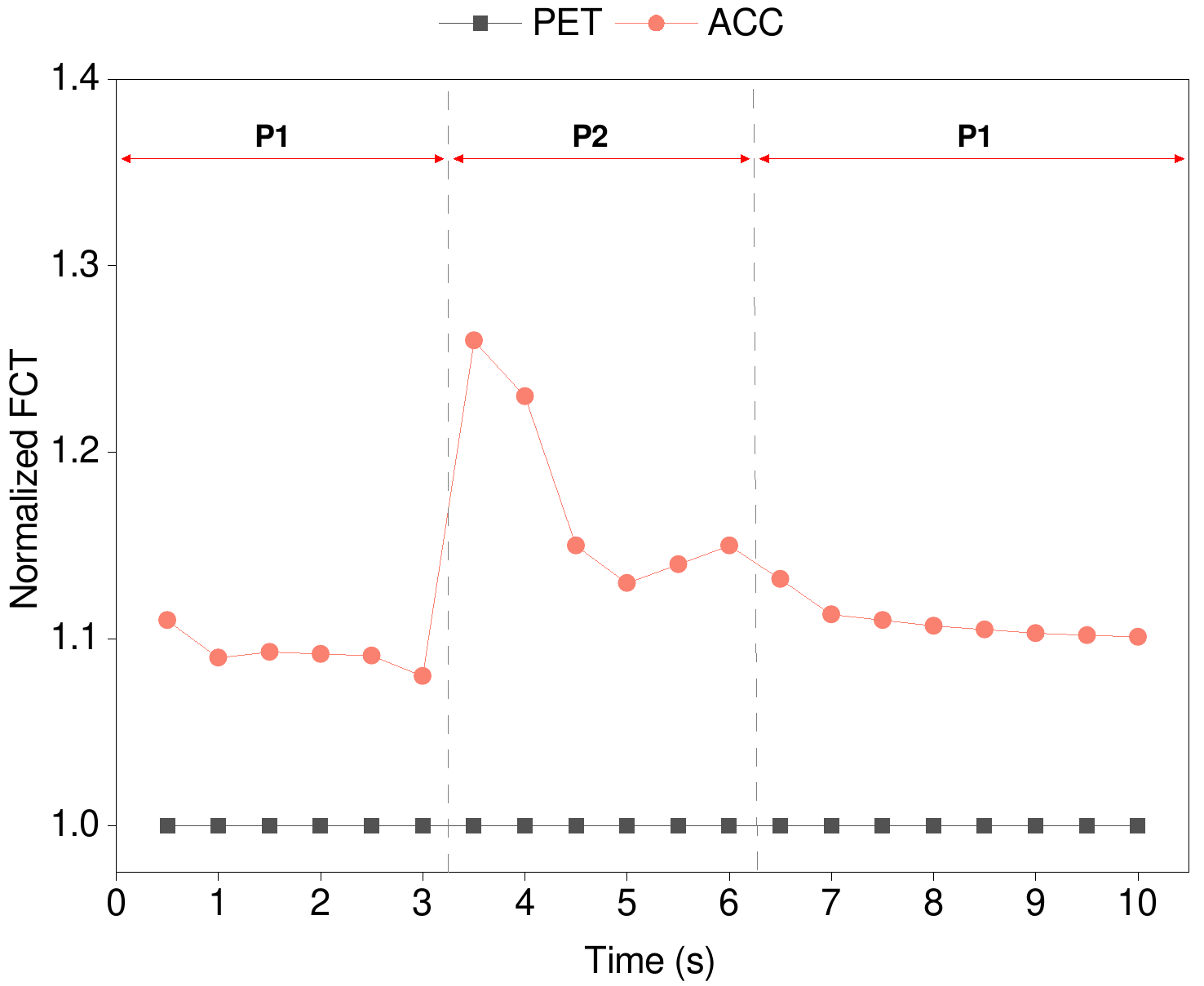}
		\caption{FCT statistics with link failure}\label{robustness}
	\end{center}
	\vspace{-0.5cm}
\end{figure}

\begin{figure}
	\begin{center}
		\includegraphics[width=5.5cm]{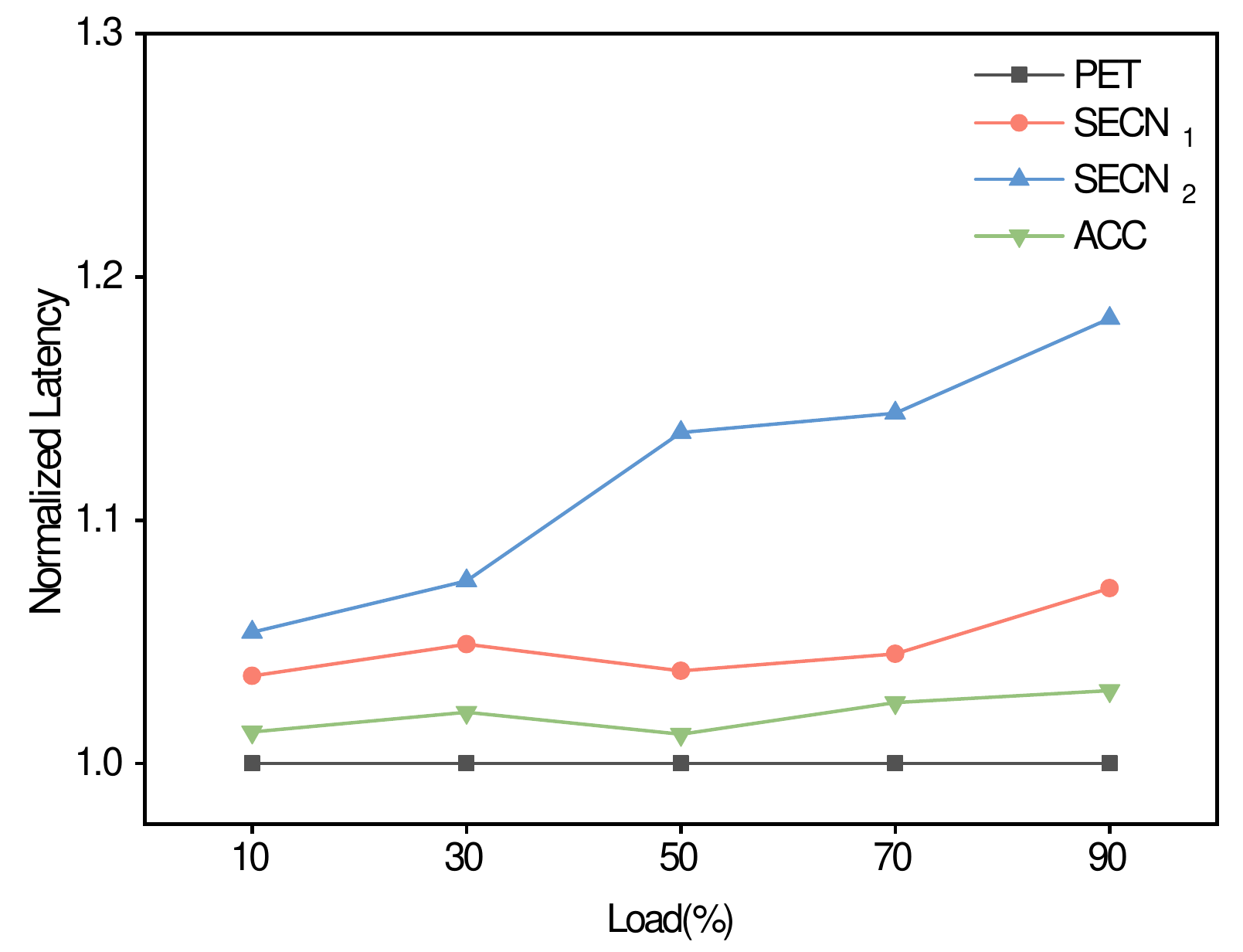}
		\caption{Latency statistics with Web Search workload}\label{latency}
	\end{center}
	\vspace{-0.5cm}
\end{figure}

\begin{figure}
	\begin{center}
		\includegraphics[width=5.5cm]{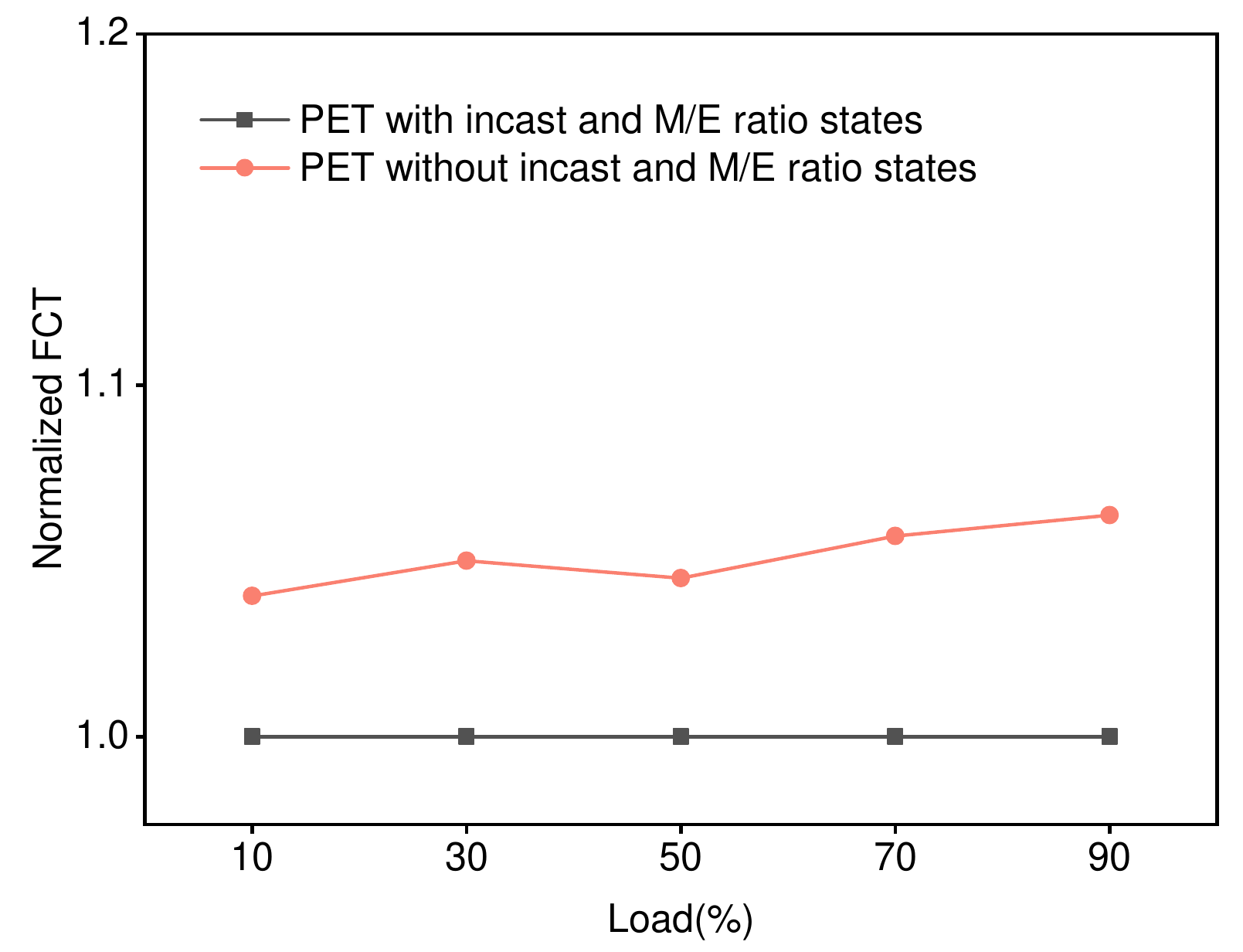}
		\caption{FCT statistics with/without incast and M/E ratio states}\label{validation}
	\end{center}
	\vspace{-0.3cm}
\end{figure}
\subsubsection{Robustness}
We evaluated the robustness of PET and ACC in the presence of link failures. 
Specifically, we randomly disconnected 10\% of switch links at 3.1s and restored these links at 6.1s. 
Fig. \ref{robustness} demonstrates that, compared to ACC, PET achieves up to 26\% reduction in the average FCT when experiencing link failures.
This reveals that PET is more robust to network failures than ACC, and can make more timely adjustments to quickly adapt to link failures resulting in a lower FCT. This is mainly due to the fact that we considered the ratio of elephant and mice flow in PET, which can help the agent react faster to traffic changes.

\subsubsection{Latency}
Latency is a crucial indicator for measuring the performance of data center networks, particularly in scenarios where latency-sensitive short flows, such as Web Search, are predominant. In light of this, we conducted a series of experiments to evaluate the latency performance of PET in the Web Search scenario. Fig. \ref{latency} illustrates the results of our evaluation, revealing that PET consistently achieves the best performance with the lowest latency across all cases in the Web Search scenario.
Specifically, PET reduces latency by up to 3\%, 7.2\% and 18.3\% compared with ACC, $\text{SECN}_{1}$ and $\text{SECN}_{2}$, respectively. This convinces that incorporating the extent of incast and the ratio of mice and elephant flows as states enables PET to effectively handle incast problems and address the queue build-up issues caused by large flows when both large and small flows coexist in the network, resulting in lower latency.

\subsubsection{Validation of States}
As introduced in Section 5, the extent of incast and the ratio of mice and elephant flows play vital roles as significant contributing factors in PET's state space. To comprehensively evaluate the actual contributions of these two factors, we further conducted some ablation experiments to compare the performance of PET with and without the inclusion of the extent of incast and the ratio of mice and elephant (M/E ratio) flows. Fig. \ref{validation}  illustrates the comparison results of the ablation experiments in terms of FCT using the Web Search workload under various network loads. 
It can be observed that  incorporating incast degree and the M/E ratio as states in PET leads to a significant reduction in the overall average FCT by up to 6.3\%. This evidently proves
the substantial contributions made by these two variables, incorporated in PET's state space, in enhancing its effectiveness in managing incast scenarios and catering to the diverse requirements of different traffic patterns.


\begin{table}[]
	\centering
	\caption{Queue Length Statistics at 60\% Load} \label{queuelen}
	\begin{tabular}{|l|l|l|}
		\hline
		\textbf{Queue Length}&
		\textbf{PET} & 
		\textbf{ACC} \\
	
		\hline
		\textbf{Average}  & 5.3KB & 6.1KB \\
		\hline
		\textbf{Variance}  & 10.2KB & 14.1KB \\
    	\hline
	\end{tabular}
	\vspace{-0.5cm}
\end{table}

\subsubsection{Discussions}
Comprehensive experiments have convinced the superior performance of our PET approach compared with both static schemes and the learning-based automatic scheme.
The major reasons behind this achievement can be briefly summarized as follows, but not limited to:
\begin{enumerate}[i)]
    \item In addition to the queue length, the current ECN threshold, and two kinds of data rates, PET also takes the incast degree and the ratio of mice and elephant flows into account when adjusting the ECN threshold, which supremely helps the RL algorithm achieve a better understanding of the network congestion tendency and differentiated requirements of different types of traffic. This evidentially benefits the algorithm in achieving low FCT in various scenarios.
    \item PET adopts the DTDE architecture, which eliminates the time cost and bandwidth cost caused by the inter-agent information exchange, enabling the algorithm more responsive to traffic pattern switching. Furthermore, DTDE framework allows each distributed agent to concentrate on its own states, which yields better robustness and a faster convergence rate.
\end{enumerate}
\section{Conclusion}
\label{conclusion}
To deal with the problems existing in current ECN-based CC schemes, this paper proposed a multi-agent IPPO-based automatic ECN tuning scheme, named PET, for high-speed data center networks. To learn a more effective tuning policy with better adaptability to network dynamics, PET incorporates six key congestion-contributing metrics, including queue length, the output data rate for each link, the extent of incast, the ratio of mice and elephant flows, the output rate of ECN marked packets, and current ECN threshold, into its multi-agent IPPO learning model. Combined with offline training and online incremental training, PET adopts the DTDE paradigm aiming to achieve better training efficiency while avoiding bandwidth overhead. Besides, PET employs the more effective multi-agent IPPO as the learning algorithm that does not require global experience replay, which can avoid unnecessary memory overhead and bandwidth costs. Experimental results prove that PET significantly outperforms both static ECN schemes (i.e., DCQCN and HPCC) and the learning-based automatic ECN tuning scheme (i.e., ACC) under various network conditions, achieving smaller flow completion time, better model stability, faster convergence rate, and higher system robustness.


%





\ifCLASSOPTIONcaptionsoff
  \newpage
\fi



%

\bibliographystyle{unsrt}
\bibliography{ref}

\begin{IEEEbiography}[{\includegraphics[width=2cm]{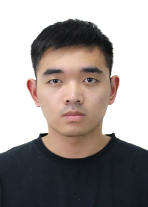}}]{Kai Cheng} received his B.S. degree in network engineering from Zhejiang University of Technology, China, in 2021. Currently, he is pursuing his MS degree at Software Engineering Institute, East China Normal University, China. His research interests include data center network, reinforcement learning, distributed computing, and machine learning.
\end{IEEEbiography}

\begin{IEEEbiography}[{\includegraphics[width=2cm]{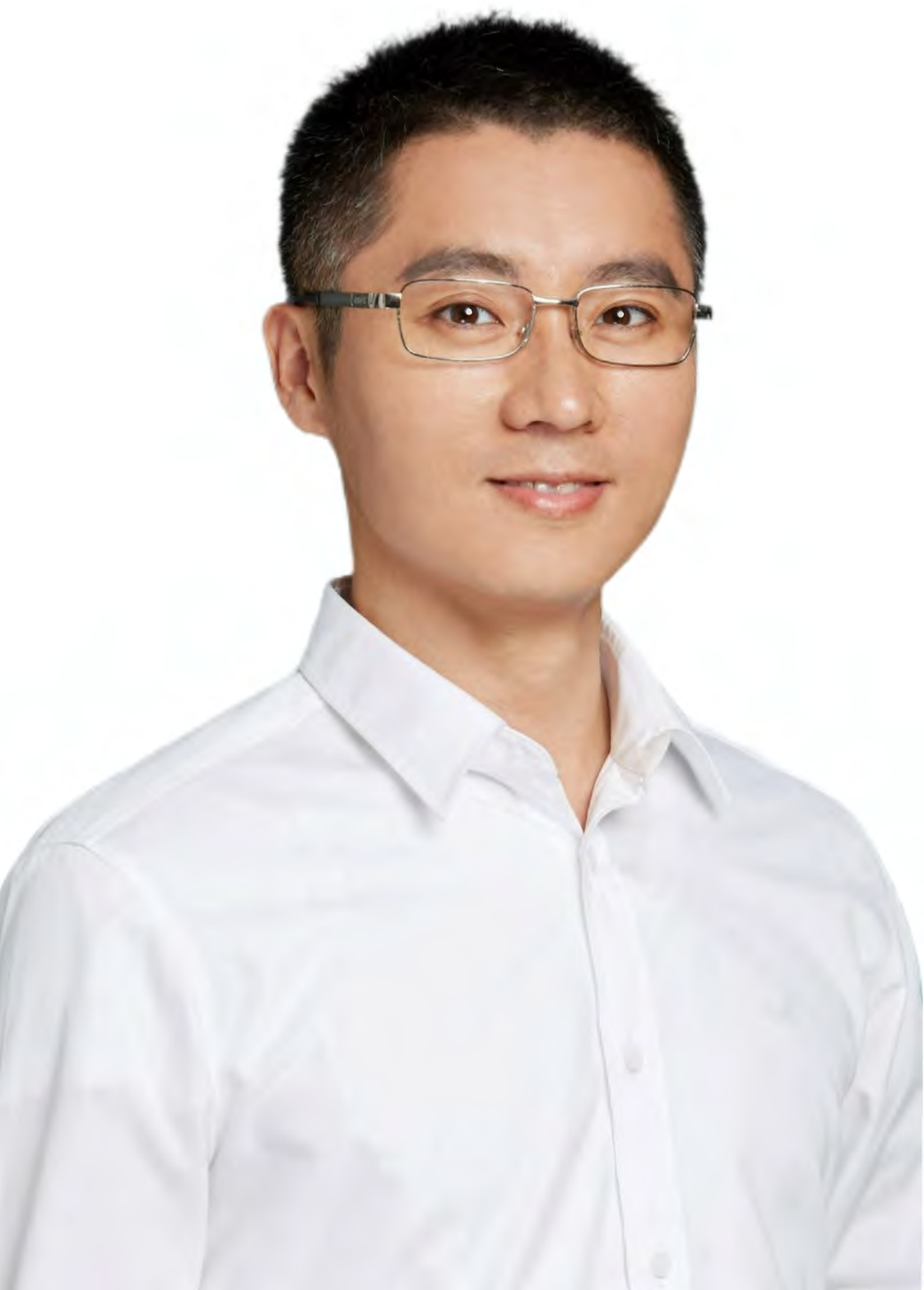}}]{Ting Wang} received the Ph.D. degree in Computer Science and Engineering from Hong Kong University of Science and Technology, Hong Kong, China, in 2015. He is currently an associate professor with the Software Engineering Institute, East China Normal University, Shanghai, China. Prior to joining ECNU in 2020, he worked at Bell Labs as a research scientist from 2015 to 2016, and at Huawei as a senior engineer from 2016 to 2020. 
His research interests include cloud/edge computing, data center networks, machine learning, and AI-aided intelligent networking.
\end{IEEEbiography}

\begin{IEEEbiography}[{\includegraphics[width=2cm]{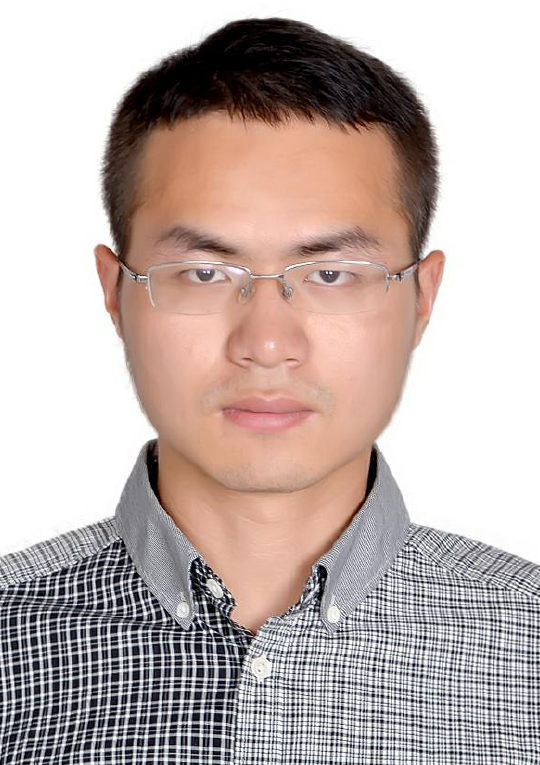}}]{Xiao Du}
received the B.S. degree in  Electronic Information Engineering from Leshan Normal University, China, in 2015, and the M.Eng. degree in Software Engineering from East China Normal University, China, in 2022. Currently, he is pursuing his Ph.D. degree at the Software Engineering Institute, East China Normal University, China. His research interests include multi-agent reinforcement learning, distributed computing, and machine learning.
\end{IEEEbiography}

\begin{IEEEbiography}[{\includegraphics[width=2.2cm]{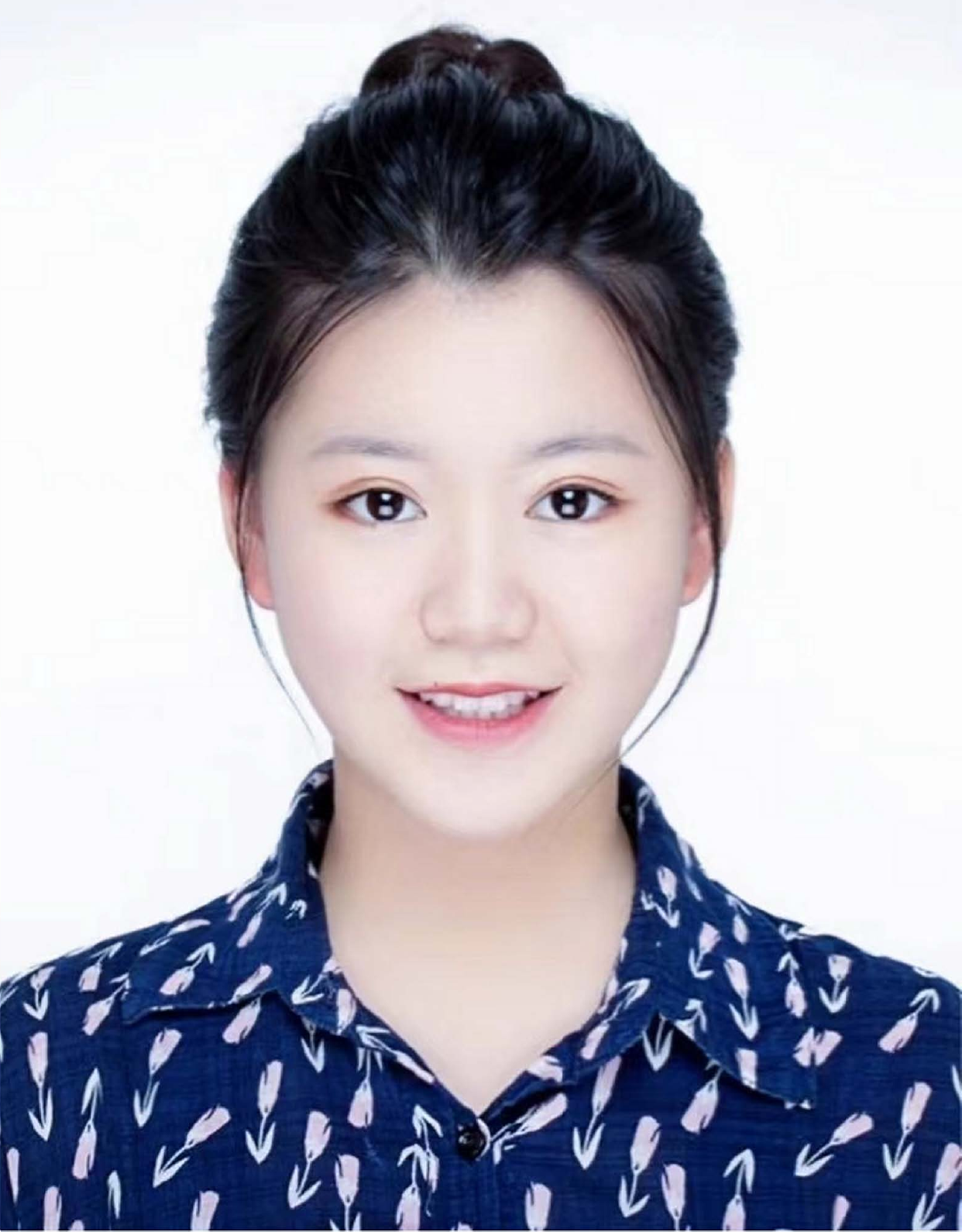}}]{Shuyi Du}
 is currently studying for a B.Eng degree at the Houston International Institute, Dalian Maritime University, China. Her research interests include machine learning, deep learning and distributed computing.
\end{IEEEbiography}

\begin{IEEEbiography}[{\includegraphics[width=2cm]{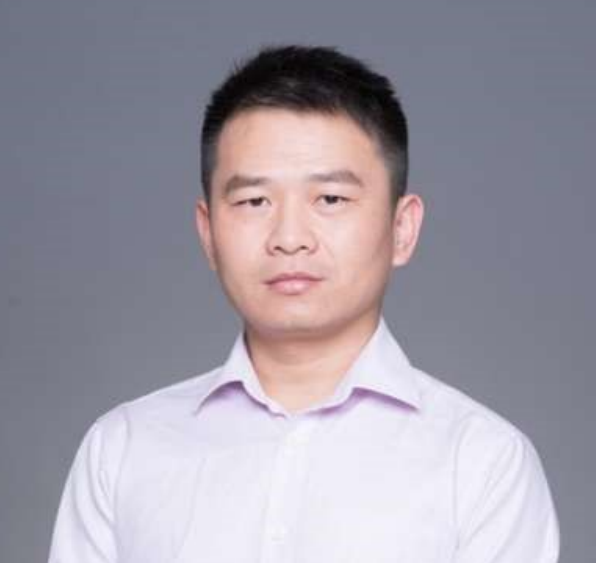}}]{Haibin Cai}
    received the B.Eng. and M.S. degrees from the National University of Defense Technology, China, in 1997 and 2004, respectively, and the Ph.D. degree from the Donghua University, Shanghai, China, in 2008. He is currently a professor with the East China Normal University. His research interests include trusted AI, intelligent perception, distributed computing and trustworthy computing.
\end{IEEEbiography}

\end{document}